%% file: ratio_paper.tex
\documentclass[aps,preprint,showpacs,superscriptaddress,linenumbers,nofootinbib,prd,12pt]{revtex4} 
\usepackage{rotating}
\usepackage[colorlinks]{hyperref}
\usepackage{lipsum}

\usepackage{amsmath}   
\usepackage{graphicx}  
\usepackage{xspace}
\usepackage{units}

\usepackage{hyperref}

\newcommand{\minerva}{MINERvA\xspace}

\newcommand{\lownu}{low-$\nu$~}

\newcommand{\ratio}{\ensuremath{R_{CC}}}
\newcommand{\nunot}{\ensuremath{\nu_0}}
\newcommand{\flux}{\ensuremath{\Phi^{\nu(\bar{\nu})}(E)}}

\newcommand{\ncc}{\ensuremath{N_\mathrm{CC}^{\nu(\bar{\nu})}(E)}}
\newcommand{\gcc}{\ensuremath{\Gamma_\mathrm{CC}^{\nu(\bar{\nu})}(E)}}

\newcommand{\bcc}{\ensuremath{B_\mathrm{CC}^{\nu(\bar{\nu})}(E)}}
\newcommand{\fsam}{\ensuremath{F^{\nu(\bar{\nu})}(E)}}
\newcommand{\aflux}{\ensuremath{A_{\Phi}^{\nu(\bar{\nu})}(E)}}
\newcommand{\afluxs}{\ensuremath{A_{\Phi}^{\nu(\bar{\nu})}(E)}}
\newcommand{\bflux}{\ensuremath{B_{\Phi}^{\nu(\bar{\nu})}(E)}}
\newcommand{\nucor}{\ensuremath{S^{\nu(\bar{\nu})}(\nu_0,E)}}

\newcommand{\fsamnu}{\ensuremath{F^{\nu}(E)}}

\newcommand{\bfluxnu}{\ensuremath{B_{\Phi}^{\nu}(E)}}
\newcommand{\nucornu}{\ensuremath{S^{\nu}(\nu_0,E)}}

\newcommand{\fsamnubar}{\ensuremath{F^{\bar{\nu}}(E)}}

\newcommand{\bfluxnubar}{\ensuremath{B_{\Phi}^{\bar{\nu}}(E)}}
\newcommand{\nucornubar}{\ensuremath{S^{\bar{\nu}}(\nu_0,E)}}

\newcommand{\realhnu}{\ensuremath{H^{\nu}(\nu_0)}}
\newcommand{\realhnubar}{\ensuremath{H^{\bar{\nu}}(\nu_0)}}

\newcommand{\xsec}{\ensuremath{\sigma_{CC}^{\nu(\bar{\nu})}(E)}}
\newcommand{\xsecnu}{\ensuremath{\sigma_{CC}^{\nu}(E)}}
\newcommand{\xsecnubar}{\ensuremath{\sigma_{CC}^{\bar{\nu}}(E)}}

\newcommand{\PRLsupp}{0}   
\ifnum\PRLsupp=0

\else

\fi

\newif\ifpdf
\ifx\pdfoutput\undefined
   \pdffalse
\else
   \pdfoutput=1
   \pdftrue
\fi
\ifpdf
   \usepackage{graphicx}
   \usepackage{epstopdf}
\else
   \usepackage{graphicx}
\fi

\begin{document}
\preprint{FERMILAB-PUB-16-595-ND}
\title{Measurement of the antineutrino to neutrino charged-current interaction cross section ratio in MINERvA}
\input{author}

\begin{abstract}
We present measurements of the neutrino and antineutrino total charged-current 
cross sections on carbon and their ratio 
using the MINERvA scintillator-tracker. The measurements 
span the energy range 2-22~GeV and were performed using 
forward and reversed horn focusing modes of the
Fermilab low-energy NuMI 
beam to obtain large neutrino and antineutrino samples.
The flux is obtained using a sub-sample of 
charged-current events at low hadronic energy transfer along 
with precise higher energy external neutrino cross section data overlapping with our energy range between 12-22 GeV. 
We also report on the antineutrino-neutrino cross section 
ratio, \ratio{},
which does not rely on external normalization information.
Our ratio measurement, obtained within the same experiment using the same technique, benefits from the cancellation of 
common sample systematic uncertainties and reaches a precision of
$\sim$5\% at low energy. Our results for the antineutrino-nucleus scattering cross section and for \ratio{} are the most precise to date 
in the energy range $E_{\nu} < 6$~GeV.

\end{abstract}

\maketitle

\section{Introduction}

Long-baseline oscillation experiments~\cite{Acciarri:2015uup}~\cite{Abe:2015zbg}, which aim to precisely measure neutrino oscillation parameters 
and constrain CP violation, will make use of neutrino and antineutrino beams 
in the few-GeV neutrino energy ($E_\nu$) range.
For appropriate baselines and energies, neutrino oscillation phenomena produce distinct shape signatures on 
either  $\nu_\mu \rightarrow \nu_e$  or  $\overline{\nu}_\mu \rightarrow \overline{\nu}_e$ appearance probabilities, which,
in matter, depend on the CP violating phase ($\delta_{CP}$) and the (unknown) sign of the mass splitting term, $\Delta m^2_{31}$.
Variations of oscillation parameters over their allowed ranges produce degenerate effects on the appearance probabilities, complicating
these measurements. 
Uncertainties in poorly constrained cross section components in this energy range produce further competing shape 
effects on the measured visible energy spectra used to extract the oscillation probabilities. 
Utilizing beams of both neutrinos and antineutrinos allows a measurement of the CP asymmetry~\cite{Marciano:2006uc}, $\mathcal{A}_{CP}$, defined as,
\begin{equation}
\mathcal{A}_{CP}=\frac{P(\nu_\mu \rightarrow \nu_e) - P(\overline{\nu}_\mu \rightarrow \overline{\nu}_e)}
{P(\nu_\mu \rightarrow \nu_e) + P(\overline{\nu}_\mu \rightarrow \overline{\nu}_e)},
\label{eq:cp_asymmetry}
\end{equation}
which can be written in terms of probability ratios.
Reducing uncertainties on the cross sections, and in particular their ratio,
$\ratio=\sigma^{\overline{\nu}}/\sigma^{\nu}$, to which $\mathcal{A}_{CP}$ is primarily sensitive, is 
essential to achieving ultimate sensitivity in oscillation measurements.

The results presented here use neutrino and antineutrino events analyzed in 
the \minerva scintillator (CH) detector 
exposed to the NuMI (Neutrinos at the Main Injector) beam.
Total cross sections are extracted from selected charged-current (CC) event samples, and incident fluxes are measured  {\it in situ}
using a sub-sample of these events at low-$\nu$ ($\nu$ is the energy transfered to the hadronic system) as in our previous result~\cite{DeVan:2016rkm}. 
The ratio, \ratio{},
is obtained by forming ratios of measured event rates in the two beam modes.
Since the measurements are performed using 
the same apparatus and flux measurement technique, 
common detector and model related systematic uncertainties cancel in the 
ratio, resulting in a precise measurable quantity that can be leveraged to 
tune models and improve knowledge of interaction cross sections.

While knowledge of neutrino cross sections has recently been improved in the low-energy region, there is a dearth 
of precise antineutrino cross section measurements at low energies (below 10 GeV)~\cite{Agashe:2014kda}. 
The cross section ratio,  \ratio{},
has recently been measured by MINOS~\cite{minos} on iron 
with a precision of $\sim$7\% at  6~GeV. At lower energies, only one 
dedicated measurement~\cite{Eichten:1973cs} (on CF$_3$Br) 
has been performed, with a precision of $\sim$20\%. 
Measurements on a range of nuclear targets are needed to 
constrain nuclear dependence which currently 
contributes significantly to modeling uncertainty.
While much of the existing data is on an iron nucleus, 
this result provides data on a light nuclear target 
(carbon). We improve on the precision of both the
antineutrino cross section and \ratio{} (by nearly a factor of four)
at low energies (2-6~GeV). 

Systematic uncertainties in our measured cross sections are dominated at the 
lowest energies by the limited knowledge of cross-section model components at low hadronic energy transfer ($\lesssim$1~GeV).
The current suite of neutrino generators ~\cite{Gazizov:2004va,genie,Buss:2011mx,Autiero:2005ve,Casper:2002sd, Battistoni:2009zzb,Golan:2012wx} 
are known to be deficient in modeling nuclear effects and detailed exclusive process rates at low energy transfer. To allow our measurement to be updated with future models, we also present the 
measured rates (corrected for detector effects and backgrounds) 
with the primary model-dependent terms factorized.

We have previously reported an inclusive CC cross section measurement~\cite{DeVan:2016rkm} using
the same data sample and method to constrain the flux shape with energy. 
The results presented here use an updated cross-section model 
which has been tuned to improve agreement with our data in the low-$\nu$ 
region~\cite{Rodrigues:2015hik} as described in Sec.~\ref{sec:xsec}.
The current work also provides a precise measurement of the ratio, \ratio{}, as well as the measured
model-independent rates for re-extracting cross sections with 
alternative generator-level models.
In addition, the antineutrino flux normalization method employed here 
improves the antineutrino cross section precision by
a factor of 1.5-1.9, which for the previous result was 
dominated by the large uncertainty ($\sim$10\%) on the model-based antineutrino normalization constraint.

\section{\minerva Experiment}
\label{sec:Detector-and-beamline}

\begin{figure*}[!htb]
\begin{center}
\includegraphics[width=0.49\columnwidth]{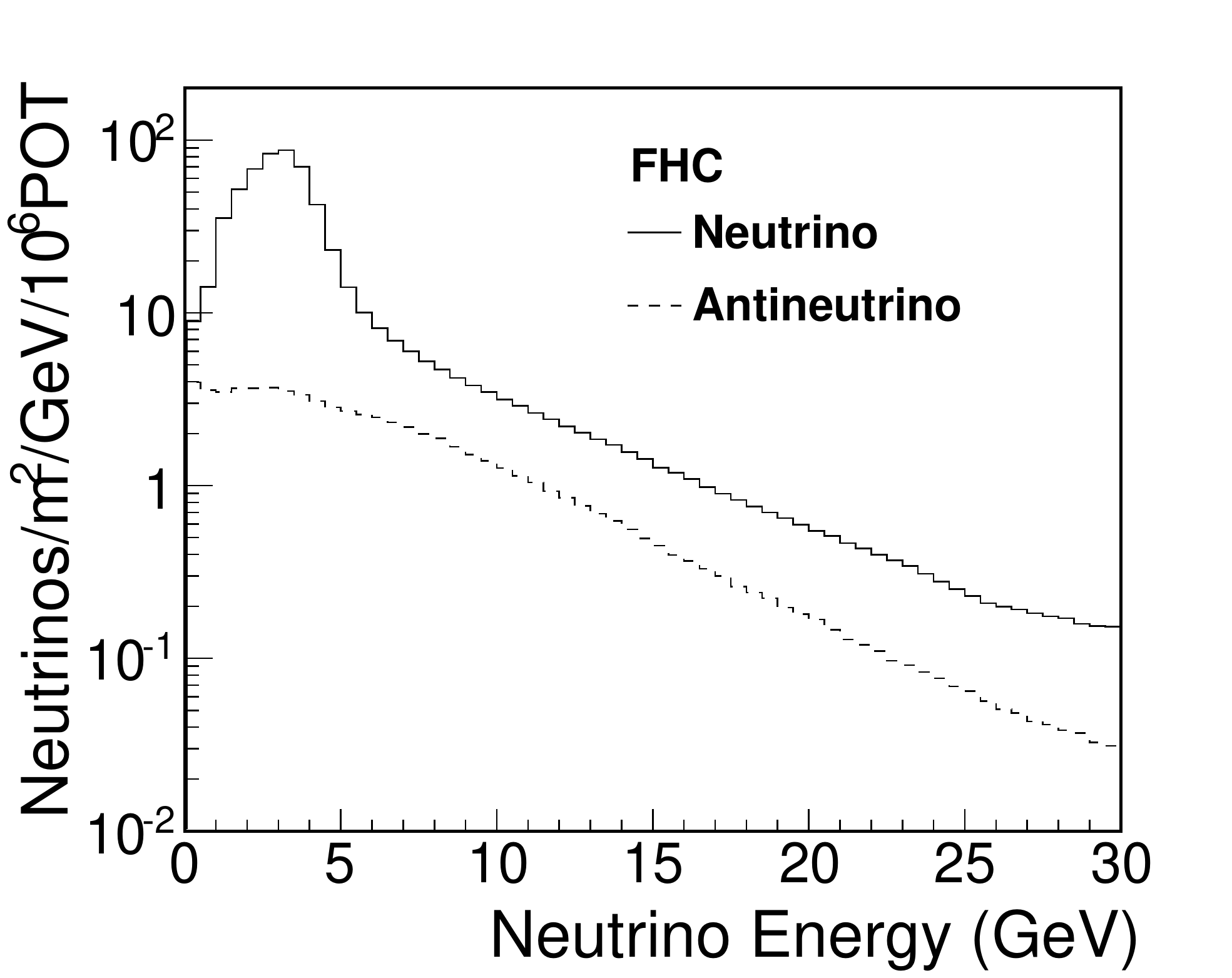}
\includegraphics[width=0.49\columnwidth]{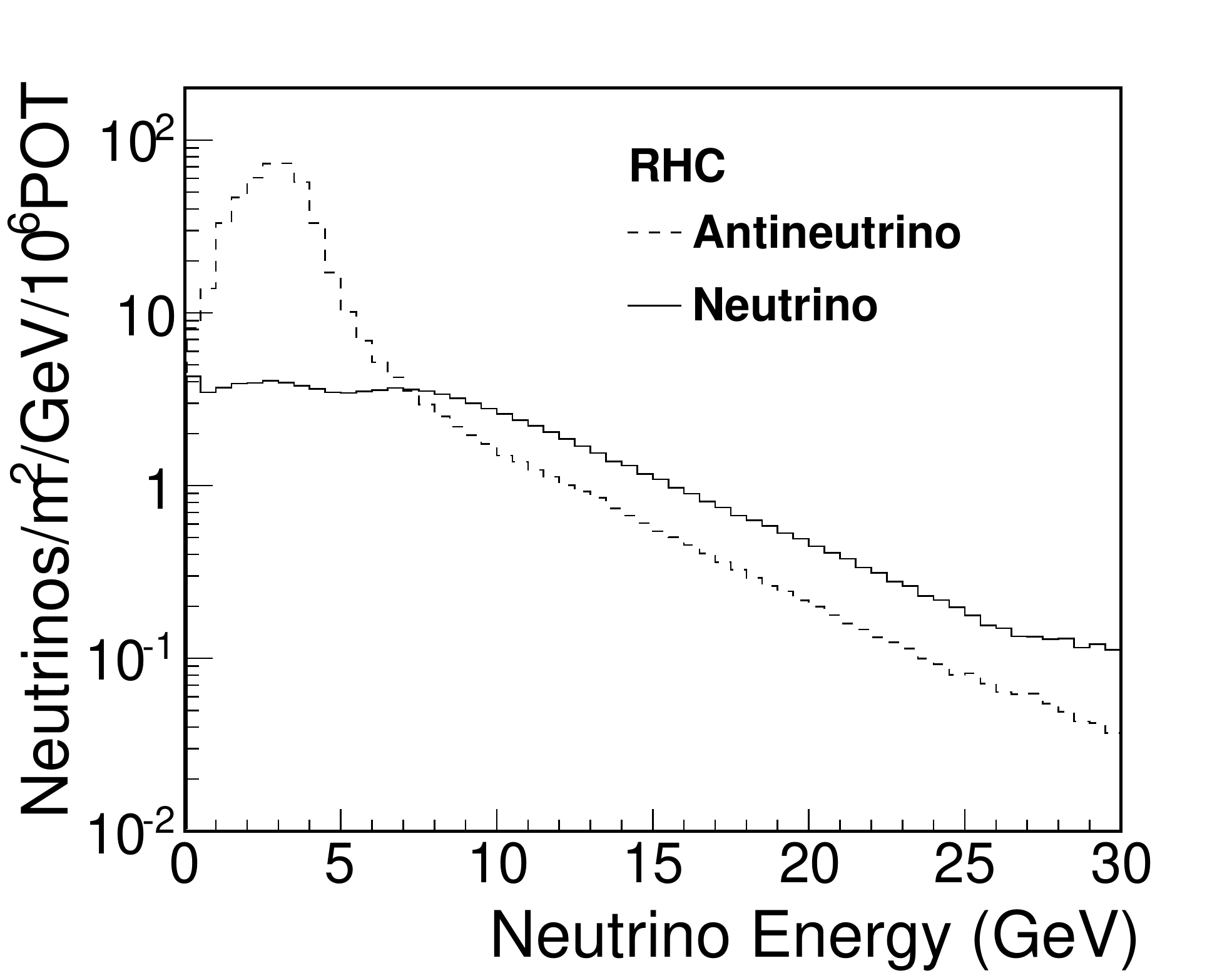}
\caption{Predicted incident neutrino fluxes at the \minerva\ detector 
in FHC (left) and RHC (right) beam modes from Ref.~\cite{Aliaga:2016oaz}. }
\label{fig:beams}
\end{center}
\end{figure*}

Muon neutrinos and antineutrinos are produced in NuMI when 120~GeV protons
from the Fermilab Main Injector strike a graphite target. Details of the NuMI beamline can be found in Ref.~\cite{numinim}.
A system of two magnetic horns is used to focus emerging secondary pions
and kaons, which are allowed to decay in the 675~m space immediately downstream of the target. 
We analyze exposures in two low-energy NuMI beam modes.
The forward horn current (FHC) mode sets the horn polarity to 
focus positively-charged secondary beam particles, which results in a 
primarily muon neutrino beam (10.4\% muon antineutrino component) with 3~GeV peak energy.
If the polarity of both horns is reversed (RHC mode) the resulting 
beam 
has a large fraction of muon antineutrinos with the same peak
beam energy and a sizable muon neutrino component (17.7\%) that 
extends to high energies.
Figure~\ref{fig:beams} shows the simulated fluxes~\cite{Aliaga:2016oaz} for muon neutrinos and antineutrinos in each mode. 
We use samples collected between March 2010 and April 2012 corresponding to 
exposures of 3.20$\times10^{20}$ protons on target (POT) 
in FHC and 1.03$\times10^{20}$ POT  in RHC beam modes. 

The \minerva fine-grained scintillator tracking detector~\cite{minervanim} is situated
approximately 1~km downstream of the NuMI target.
The active detector consists of triangular scintillator strips with height 
1.7~cm and base 3.3~cm arranged into hexagonal X, U and V planes (at 60 degrees with respect to one another) and 
giving single-plane position resolution of about 2.5~mm. 
We use events originating in the 6~ton fully-active scintillator region that is primarily 
composed of carbon nuclei (88.5\% carbon,  8.2\% hydrogen,  2.5\% oxygen and a number of other nuclei
that make up the remaining fraction, by mass). We report results on a carbon target by
correcting for the \minerva\ target proton excess (see Sec.~\ref{sec:lownuflux}).

The downstream most plane of \minerva\ is positioned 2~m 
upstream of the magnetized MINOS Near Detector~\cite{Michael:2008bc} (MINOS ND), 
which is used to contain and momentum analyze muons exiting
the \minerva active detector volume.
The detector geometry changes from sampling after every iron plane (2.54~cm thickness) to sampling every five iron-scintillator units after the first 7.2~m. This produces features in the measured muon momentum distribution and acceptance which will be discussed below.
For FHC (RHC) beam mode the MINOS ND toroidal magnetic field 
is set to focus negatively (positively) charged muons.
Measurement of the direction of track curvature is used to tag the charge-sign of tracks, which  
is crucial to reducing the large wrong-sign beam background in RHC mode.

\section{Monte Carlo Simulation}
\label{sec:xsec}

We use a custom \minerva-tuned modification of GENIE 2.8.4 \cite{Andreopoulos:2009rq, Andreopoulos:2015wxa} referred to here as ``GENIE-Hybrid''
as input to simulated event samples as well as for the model correction terms needed to obtain our
default cross section results. This model incorporates improved modeling of low-$\nu$ cross section components
and is similar to that described in Ref.~\cite{Rodrigues:2015hik}. 
GENIE 2.8.4 uses a modified version of the relativistic Fermi gas model of the nucleus, which is inadequate to precisely describe  
neutrino scattering data at low three-momentum transfer such as quasi-elastic (QE) and $\Delta$(1232) 
resonance production. For QE events, we use the Random Phase Approximation (RPA)~\cite{Nieves:2004wx} model, which includes long-range nucleon-nucleon correlations 
to more accurately characterize scattering from a nucleon bound in a nucleus.
We also include the Valencia ``2p2h" model contribution~\cite{Nieves:2011pp} of the neutrino 
interacting with a correlated nucleon pair that populates the energy transfer
region between the QE and $\Delta$-resonance events.
Since even this does not adequately cover the observed signal excess in this region~\cite{Rodrigues:2015hik}, we include additional modeling uncertainties from this contribution. 
In addition, we reduce the GENIE single pion 
non-resonant component\footnote{The corresponding GENIE parameter is $R_{bkg}^{\nu n CC 1 \pi}$ for neutrino and $R_{bkg}^{\bar{\nu} p CC 1 \pi}$ for antineutrino~\cite{Andreopoulos:2015wxa}.} with initial state $\nu+n$ (or $\bar{\nu}+p$)
by 57$\%$, which has been shown to improve agreement with observed deuterium data \cite{Rodrigues:2016xjj}.

\section{Technique Overview}

Events studied in this analysis are categorized as charged-current events 
by the presence of a long track originating from the  primary interaction vertex
which extrapolates into the MINOS ND.
The inclusive sample, \ncc{}, is the number of measured charged current events in neutrino energy bin $E$.
We define $\mathcal{R}^{\nu(\bar{\nu})}(E)$, which is related to the fiducial cross section, as
\begin{equation}
\mathcal{R}^{\nu(\bar{\nu})}(E)=\frac{(\ncc{}-\bcc{})\times A^{{\nu(\bar{\nu})},{\rm DET}}_{CC}(E)}
{(\fsam{}-\bflux{})\times\aflux{}},
\label{eq:uncorrecteddata}
\end{equation}
\noindent
where superscript $\nu$ ($\bar{\nu}$) refers to neutrino (antineutrino).  \fsam{} is the ``flux sample'' obtained from a 
subset of \ncc{} with low hadronic energy (discussed below). The terms $\bcc{}$ and $\bflux{}$ are
backgrounds due to neutral current and wrong-sign beam contamination in the inclusive and flux 
samples, respectively. Terms $A^{{\nu(\bar{\nu})},{\rm DET}}_{CC}(E)$ and $\aflux{}$ correct the cross 
section and flux respective samples for detector resolution and bin-migration effects. 
The numerator of Eq.~\eqref{eq:uncorrecteddata}, $\gcc$, 
\begin{equation}
\gcc=(\ncc{}-\bcc{})\times A^{{\nu(\bar{\nu})},{\rm DET}}_{CC}(E),
\label{eq:uncorrecteddata_inclusive}
\end{equation}
is the fiducial event rate and is tabulated below.
To obtain the incident beam flux, 
we employ the ``low-$\nu$'' method described previously~\citep{Mishra:1990ax,DeVan:2016rkm,seligman,minos}.
In brief, the differential dependence of the cross section in terms of $\nu$ is expanded in $\nu/E$ as
\begin{equation}
\frac{d\sigma^{\nu, \bar{\nu}}}{d\nu}=A\left(1 + \frac{B^{\nu, \bar{\nu}}}{A}\frac{\nu}{E}-\frac{C^{\nu, \bar{\nu}}}{A}\frac{\nu^{2}}{2E^{2}}\right),
\label{eq:dsigmadnu}
\end{equation}
where $E$ is the incident neutrino energy. The coefficients $A$, $B^{\nu, \bar{\nu}}$, and $C^{\nu, \bar{\nu}}$ depend on integrals over structure functions
(or form factors, in the low energy limit). 
\begin{equation}
A=\frac{G^{2}_{F}M}{\pi}\int F_{2}(x)dx,
\label{eq:aterm}
\end{equation}

\begin{equation}
B^{\nu, \bar{\nu}}=-\frac{G^{2}_{F}M}{\pi}\int (F_{2}(x)\mp xF_{3}(x))dx,
\label{eq:bterm}
\end{equation}
and
\begin{equation}
C^{\nu, \bar{\nu}}=B^{\nu, \bar{\nu}}-\frac{G^{2}_{F}M}{\pi}\int F_{2}(x)
\left(\frac{1+\frac{2Mx}{\nu}}{1+R_{L}}-\frac{Mx}{\nu}-1\right)dx.
\end{equation}

In the limit of $\nu/E \rightarrow 0$, the $B$ and $C$ terms vanish and both cross sections 
approach $A$ (defined in Eq.~\eqref{eq:aterm}),
which is the same for neutrino and antineutrino probes scattering off an 
isoscalar target (up to a small correction for quark mixing).
We count events below a maximum $\nu$ value ($\nu_0$) and 
apply a model-based correction
\begin{equation}
S^{\nu(\bar{\nu}),\nu_0}(E) =\frac{\sigma^{\nu(\bar{\nu})}(\nu_0,E)}{\sigma^{\nu(\bar{\nu})}(\nu_0, E\to\infty)},
\label{eq:lownu}
\end{equation}
to account for $\nu/E$ and  $(\nu/E)^{2}$ terms in Eq.~\eqref{eq:dsigmadnu}. The 
numerator in Eq.~\eqref{eq:lownu} is the value of the
integrated cross section below our chosen $\nu_0$ cut at energy $E$, and the denominator 
is its value in the high energy limit.
For antineutrinos, the structure functions in Eq.~\eqref{eq:bterm} add, resulting in a larger energy dependent
correction term than for the neutrino case where they are subtracted and partially cancel.
The flux is then proportional to the corrected low-$\nu$ rate
\begin{equation}
\flux \propto \frac{(\fsam{}-\bflux{})\times\aflux{}}
{\nucor{}}.
\label{eq:prop_flux}
\end{equation}

We obtain a quantity that is proportional to the total CC cross section,
\begin{equation}
\xsec{} \propto  \mathcal{R}^{\nu(\bar{\nu})}  \times \nucor{} \times A^{\nu(\bar{\nu}),{\rm KIN}}_{CC}(E),
\label{eq:prop_sigma_nu}
\end{equation}
by applying a correction, $A^{\nu(\bar{\nu}),{\rm KIN}}$, for regions outside of our experimental acceptance.
The term $A^{\nu(\bar{\nu}),{\rm KIN}}$ (discussed in Sec.~\ref{sec:rates}) is computed from a generator 
level Monte Carlo model.
The rates, $\mathcal{R}^{\nu}$ and $\mathcal{R}^{\bar{\nu}}$, in each beam mode
are used to obtain the ratio
\begin{widetext}
\begin{equation}
\ratio(E)=\frac{\xsecnubar{}}{\xsecnu{}}= 
\frac{\mathcal{R}^{\bar{\nu}}}{\mathcal{R}^{\nu}} 
\left( \frac{ A^{\bar{\nu}, KIN}_{CC}(E) \times \nucornubar{} \times \realhnu{} }
{A^{\nu, KIN}_{CC}(E) \times \nucornu{} \times \realhnubar{}}
\right).
\label{eq:ratio}
\end{equation}
\end{widetext}
The terms $\realhnu{}$ and $\realhnubar{}$, which supply the absolute flux normalization in 
the low-$\nu$ method for neutrinos and antineutrinos, respectively,
are related in the Standard Model and nearly cancel in this ratio.
The measurements are performed using 
the same detector and beamline, which reduces the effect of some experimental uncertainties.
The ratio measured in this technique also benefits from cancellation of 
correlated model terms; this cancellation reduces the modeling component of the systematic uncertainty 
relative to that for either neutrino or antineutrino measured cross section.

\section{Event Reconstruction and Selection}

Neutrino events are reconstructed using timing and spatial 
information of energy deposited in the \minerva  scintillator.
Hits are grouped in time into ``slices'' and within a slice, spatially into 
``clusters'' which are used along with pattern recognition to identify tracks.
The CC-inclusive event sample, denoted \ncc{}, is selected by requiring a primary 
track matched into the MINOS ND. MINOS-matched track 
momentum, $E_\mu$, is reconstructed using either range, for tracks that stop and deposit 
all of their energy in the MINOS ND, or the measured curvature of 
the trajectory, for tracks which exit the MINOS ND.
Tracks measured from range in MINOS have a momentum resolution of order 5\% while those 
measured from curvature typically have a resolution of order 10\%.
Clusters not associated with the MINOS-matched muon track form the recoil system and
are calorimetrically summed to obtain the hadronic energy, $\nu$.
Neutrino energy is constructed from the sum $E_\nu= E_\mu + \nu$. 
An event vertex is assigned by tracking the muon upstream
through the interaction region until no energy is
seen in an upstream cone around the track. The vertex is required to be within the fiducial region of the scintillator.

Additional track requirements are applied to improve energy resolution and acceptance.
The track fitting procedure in the MINOS spectrometer yields a measurement of the 
momentum
with an associated fractional uncertainty, which is required to be less than 30\%.  
The charge-sign is determined by measuring the track curvature and is required to be negative for tracks in FHC mode and positive for those in RHC mode. 
We also require the 
muon track candidate to have a minimum energy $E_{\mu}>1.8$~GeV and a maximum angle 
$\theta_\mu<0.35$~rad (20 deg) with respect to the beam direction in the lab frame. 
The portion of the track in MINOS is required to not pass through the uninstrumented coil hole region.
Events in which the muon track ends less than 80 cm from the center of the coil hole are also removed.
This removes 0.8\% (0.4\%) events from the neutrino (antineutrino) CC-inclusive sample.

The flux-extraction technique uses $\fsam{}$, the number of CC-inclusive events in an energy bin 
below a
maximum $\nu$ value. 
We choose this maximum value ($\nu_0$) to vary with energy, keeping the energy dependent
contributions in Eq.~\eqref{eq:dsigmadnu}
small ($\lesssim$0.1 for neutrinos and $\lesssim$0.2 for antineutrinos)
in the region where  modeling uncertainties are sizable ($E_\nu< 7$~GeV), while at higher energies where we normalize to external data (12-22~GeV), it is 
increased to improve statistical precision. 
The values are  $\nu_0 =$ 0.3~GeV for $E_\nu<3$~GeV, $\nu_0 =$ 0.5~GeV for $3<E_\nu<7$~GeV, 
$\nu_0 =$ 1~GeV for $7<E_\nu<12$~GeV and $\nu_0 =$ 2~GeV for $E_\nu>12$~GeV.
The inclusive and flux sample overlap is less than 50\% (60\%) for neutrinos (antineutrinos).

\subsection{Event Rates}
\label{sec:rates}

\begin{figure*}[!htb]
\begin{center}
\includegraphics[width=0.49\columnwidth]{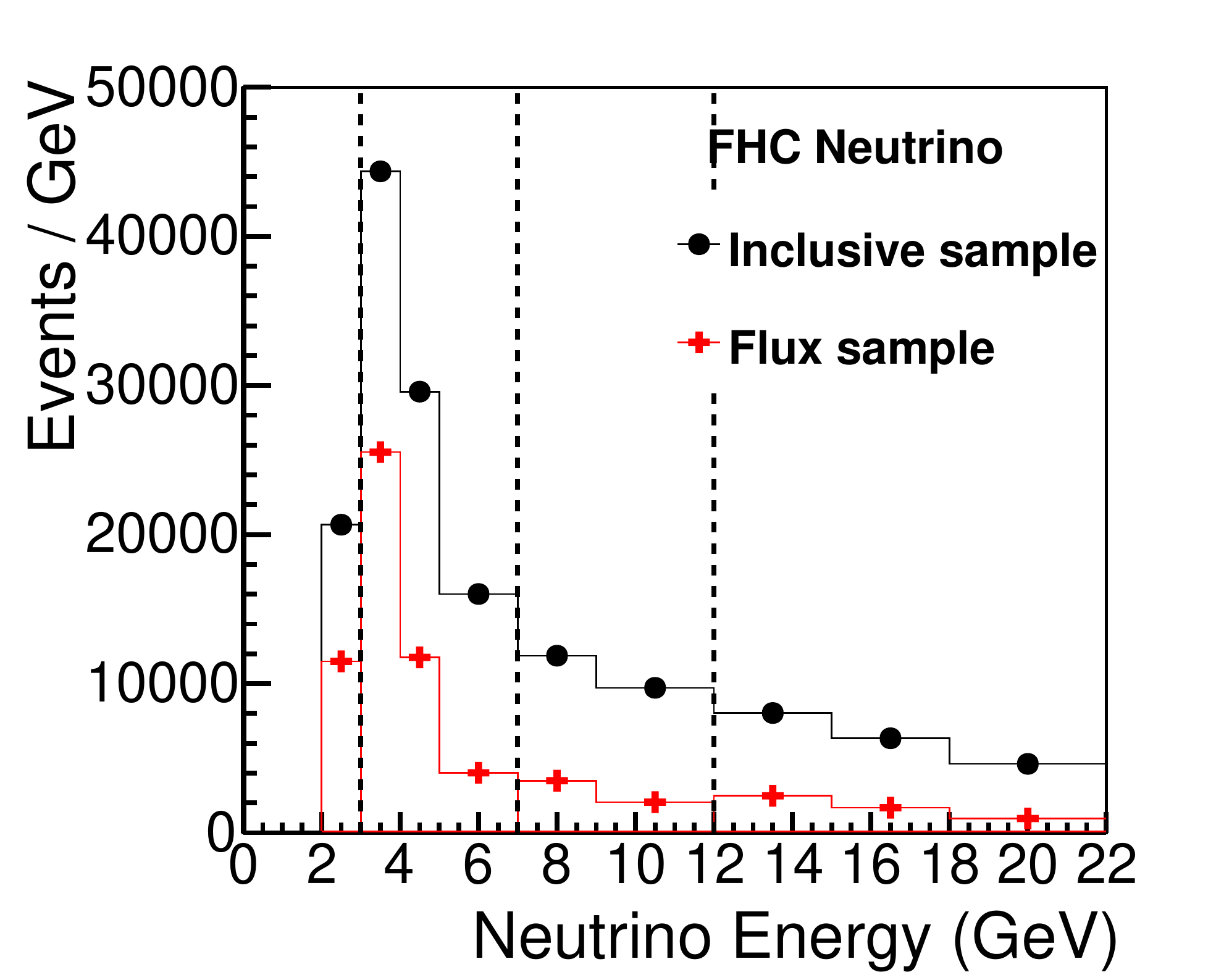}
\includegraphics[width=0.49\columnwidth]{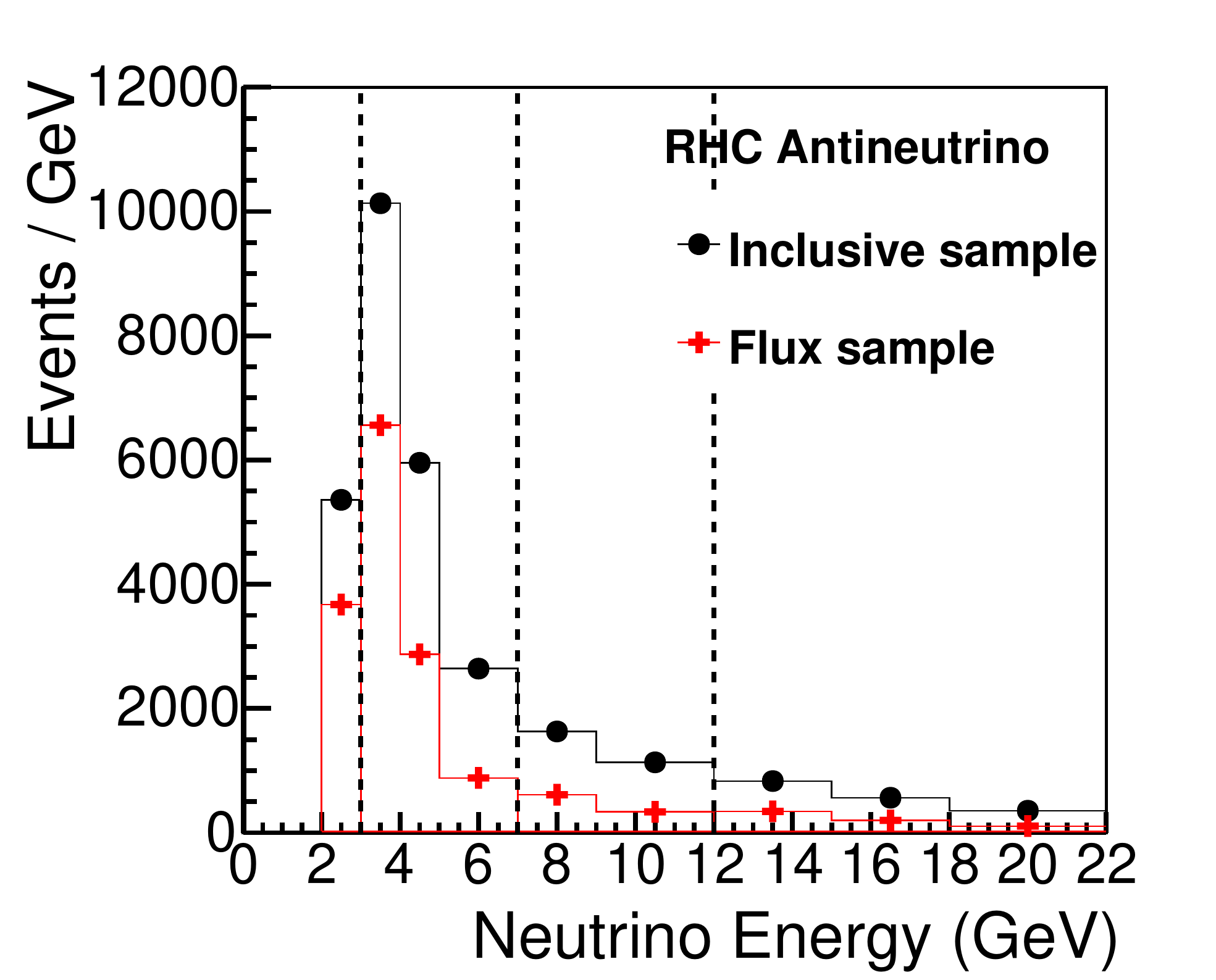}
\caption{
Neutrino inclusive ($N^{\nu(\bar{\nu})}_{CC}$) and \lownu{} flux sample ($F^{\nu(\bar{\nu})}$) yields for FHC neutrino (left) and RHC antineutrino (right) modes. The dashed lines are plotted at the values where the flux sample $\nunot$ is changed. Statistical errors are too small to be visible on the points.}
\label{fig:samples}
\end{center}
\end{figure*}

Figure~\ref{fig:samples} shows the measured inclusive  and flux sample rates in the two beam modes. 
The fiducial event rate, $\gcc$, (Eq.~\eqref{eq:uncorrecteddata_inclusive})
is determined by  removing sample backgrounds and 
applying corrections for experimental acceptance. The components are 
described below and tabulated in 
Table~\ref{tab:rawdata_nu}、.

Backgrounds are dominated by the contribution from tracks with misidentified charge-sign which arise
from the  wrong-sign beam flux component (wrong-sign contamination). The background peaks
at high energies in RHC mode (about 4\% above 10 GeV in the inclusive sample). 
The charge-sign and track quality requirements effectively reduce the wrong-sign contamination.
The remaining background is estimated
using the simulated wrong-sign beam flux shown in  Fig.~\ref{fig:beams}.
The neutral current contribution is negligible ($\ll 1$\%) in both beam modes.
\begin{table*}[t]
\begin{tabular}{|c |c|c c c c c c|c c c c c c|}
\hline
\nunot(GeV)&$E$(GeV) & $N^{\nu}_{CC}$  & $B^{\nu}_{CC}$ &  $A_{CC}^{\nu,\rm DET}$ &  $F^{\nu}$&  $B^{\nu}_{\Phi}$  & $A^{\nu}_{\phi}$ & $N^{\bar{\nu}}_{CC}$ & $B^{\bar{\nu}}_{CC}$ &  $A_{CC}^{\bar{\nu},\rm DET}$ &  $F^{\bar{\nu}}$& $B^{\bar{\nu}}_{\Phi}$  & $A^{\bar{\nu}}_{\phi}$ \\
\hline
0.3&2-3&20660 &53 &2.38& 11493 &29 &1.94&5359 &18 &1.99& 3673 &6 &1.60\\\hline
&3-4&44360 &61 &2.30& 25530  &19 &1.76&10133 &25 &1.94& 6560&4 &1.56\\
0.5&4-5&29586 &65 &1.92& 11765 &13 &1.45&5955 &24 &1.65& 2871&2 &1.36\\
&5-7&32026 &170 &1.70& 8046  &29 &1.34&5284 &74 &1.47& 1764 &4 &1.27\\\hline
1.0&7-9&23750 &171 &1.86& 6980  &32 &1.59&3261 &102 &1.58& 1224 &6 &1.50\\
&9-12&29161 &207 &1.95& 6165 &31 &1.60&3400 &141 &1.66& 1007 &9 &1.53\\\hline
&12-15&24093 &158 &1.94& 7438  &39 &1.42&2496 &115 &1.63& 1033&9 &1.42\\
2.0&15-18&19011 &104 &1.85& 5041 &17 &1.28&1690&77 &1.48& 595&6 &1.23\\
&18-22&18475 &98 &1.78& 3826 &14 &1.25&1418 &72 &1.44& 427&5 &1.23\\
\hline
\end{tabular}
\caption{Neutrino and antineutrino inclusive,  $N^{\nu(\bar{\nu})}_{CC}$,  and flux sample,  $F^{\nu(\bar{\nu})}$, yields along with 
corresponding background contributions ($B^{\nu(\bar{\nu})}_{CC}$ and $B^{\nu(\bar{\nu})}_{\Phi}$, respectively). The 
acceptance term, $A_{CC}^{\nu(\bar{\nu}),\rm DET}$, is applied to obtain the fiducial event rate,
$\gcc$,  from Eq.~\eqref{eq:uncorrecteddata_inclusive}.}
\label{tab:rawdata_nu}
\end{table*}

We correct for the experimental acceptance effects using a full  
detector simulation along with a tuned version of 
GENIE Monte Carlo (GENIE-Hybrid) which is described in Sec.~\ref{sec:xsec}. 
We separate experimental acceptance terms into two contributions. 
The term $A_{CC}^{\nu(\bar{\nu}),\rm DET}$,  which represents the ratio of the number of events generated in a given neutrino energy bin to the number 
reconstructed in our event sample, accounts for detector resolution smearing and bin migration effects. 
Final state interaction (FSI) effects, which arise from reinteractions of emerging 
final state particles in the target nucleus, change the measured hadronic energy
and also affect $A_{CC}^{\nu(\bar{\nu}),\rm DET}$. This bin migration effect is included in our Monte Carlo simulation
model. The term \afluxs{} is defined similarly with an additional maximum $\nu$ requirement.
%
The fiducial event rate 
depends only on $A_{CC}^{\nu(\bar{\nu}),\rm DET}$ and \afluxs{}
and 
is nearly generator model independent.
The kinematic acceptance, $A^{KIN}_{CC}$,
defined as the ratio of all generated events 
in a given bin to those with muon energy $E_\mu>$1.8~GeV and angle $\theta_\mu<0.35$~rad,
must be applied to obtain a total cross section from the fiducial event rate.
This term is computed directly from a generator level model. It is tabulated for our default model
along with other model-dependent corrections
in Table~\ref{tab:corrections_GENIE_nu}.
Nearly all muons in the selected flux sample automatically pass the kinematic cuts (except for 
a small fraction in the first energy bin which is computed to be 5.1\% using the GENIE-Hybrid model and 4.9\% using NuWro~\cite{Golan:2012wx}).
We therefore only report one acceptance, $A_{\Phi}$, which includes the kinematic
contribution in the flux sample.

\begin{figure*}[!htb]
\begin{center}
\includegraphics[width=0.49\columnwidth]{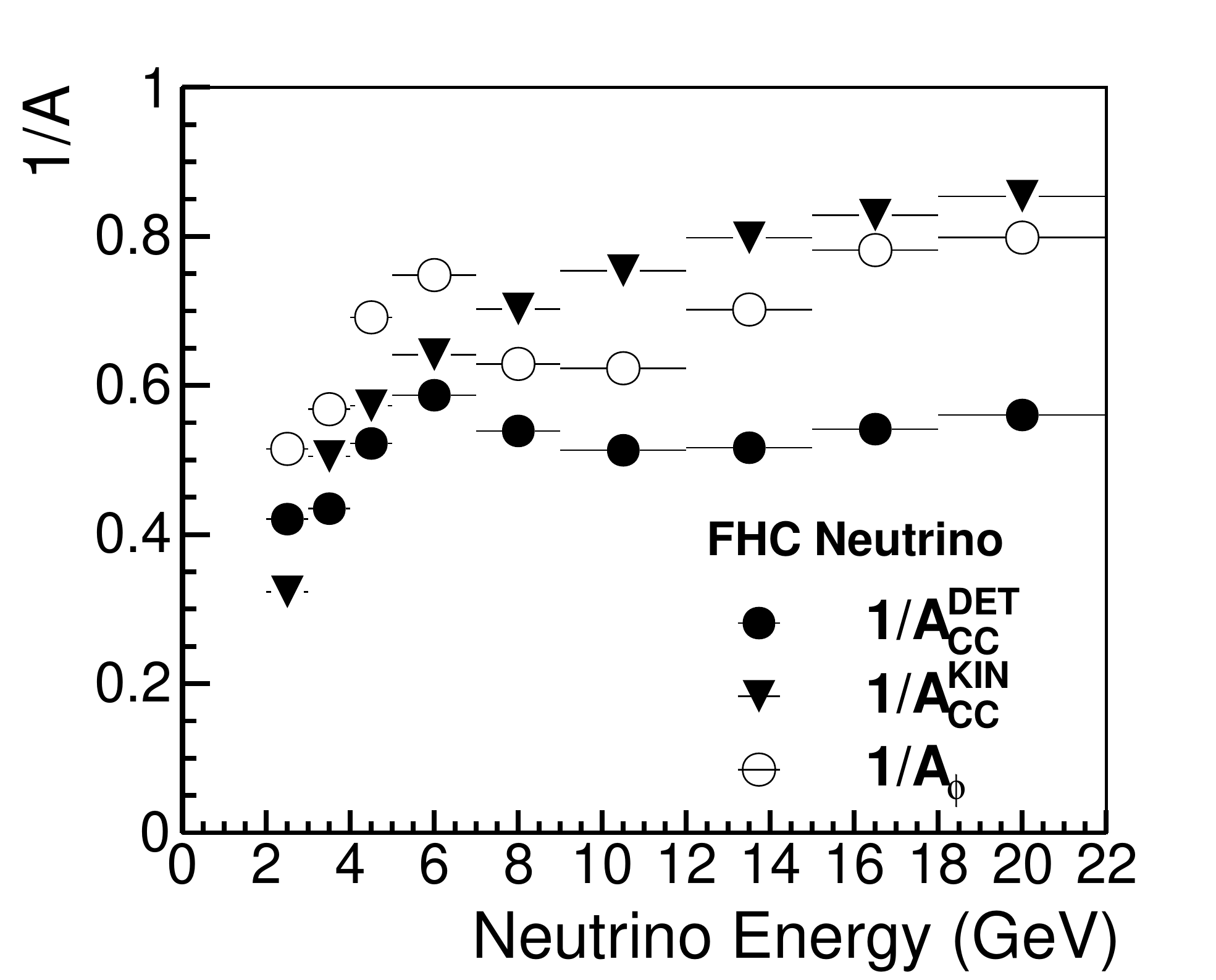}
\includegraphics[width=0.49\columnwidth]{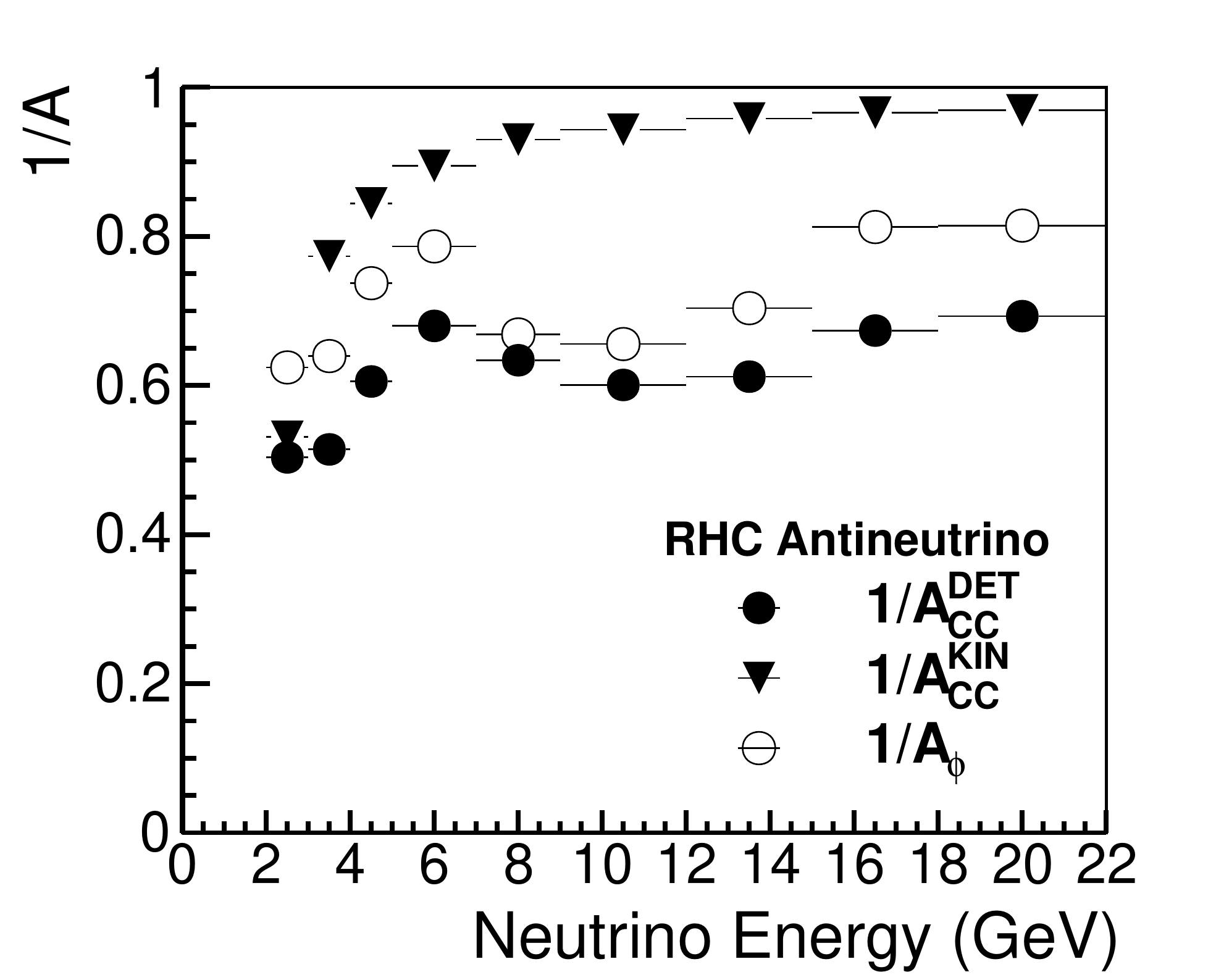}
\caption{Reciprocal of acceptance components ($1/A_{CC}^{\rm DET}$,$1/A^{\rm KIN}$) for cross section and ($1/A_{\Phi}$) for flux samples of FHC neutrinos (left) and RHC antineutrinos (right). }
\label{fig:acceptance}
\end{center}
\end{figure*}

Figure~\ref{fig:acceptance} shows the size of the acceptance correction terms for each sample.
Kinematic acceptance is most important at lowest energies 
(primarily below 3~GeV), which have the largest fraction of
events below muon energy threshold. The kinematic thresholds result in 
poorer overall acceptance at all energies for neutrinos compared with antineutrinos. This is a
consequence of the different inelasticity ($y=\nu/E_\nu$)
dependence of the two cross sections, which produce a 
harder muon energy distribution for antineutrinos with correspondingly more forward-going muons.  The flux sample with the $\nu<\nu_0$ requirement also selects a harder muon spectrum  and 
results in better corresponding acceptance relative to the inclusive sample in both modes.
The detector acceptance is above 50\% for neutrino energies greater than 5~GeV. 
The shapes of  $1/A_{CC}^{\rm DET}$ and $1/A_{\Phi}$ are affected by the MINOS ND sampling geometry as well as the two methods of measuring momentum (from range and from curvature),  which have different resolution. The dip in the 6-10~GeV region results from the contained (range) momentum sample decreasing while the curvature sample, which has poorer resolution, is becoming dominant. 

\section{Low-$\nu$ Flux Extraction}
\label{sec:lownuflux}

\begin{figure*}[!htb]
\begin{center}
\includegraphics[width=0.49\columnwidth]{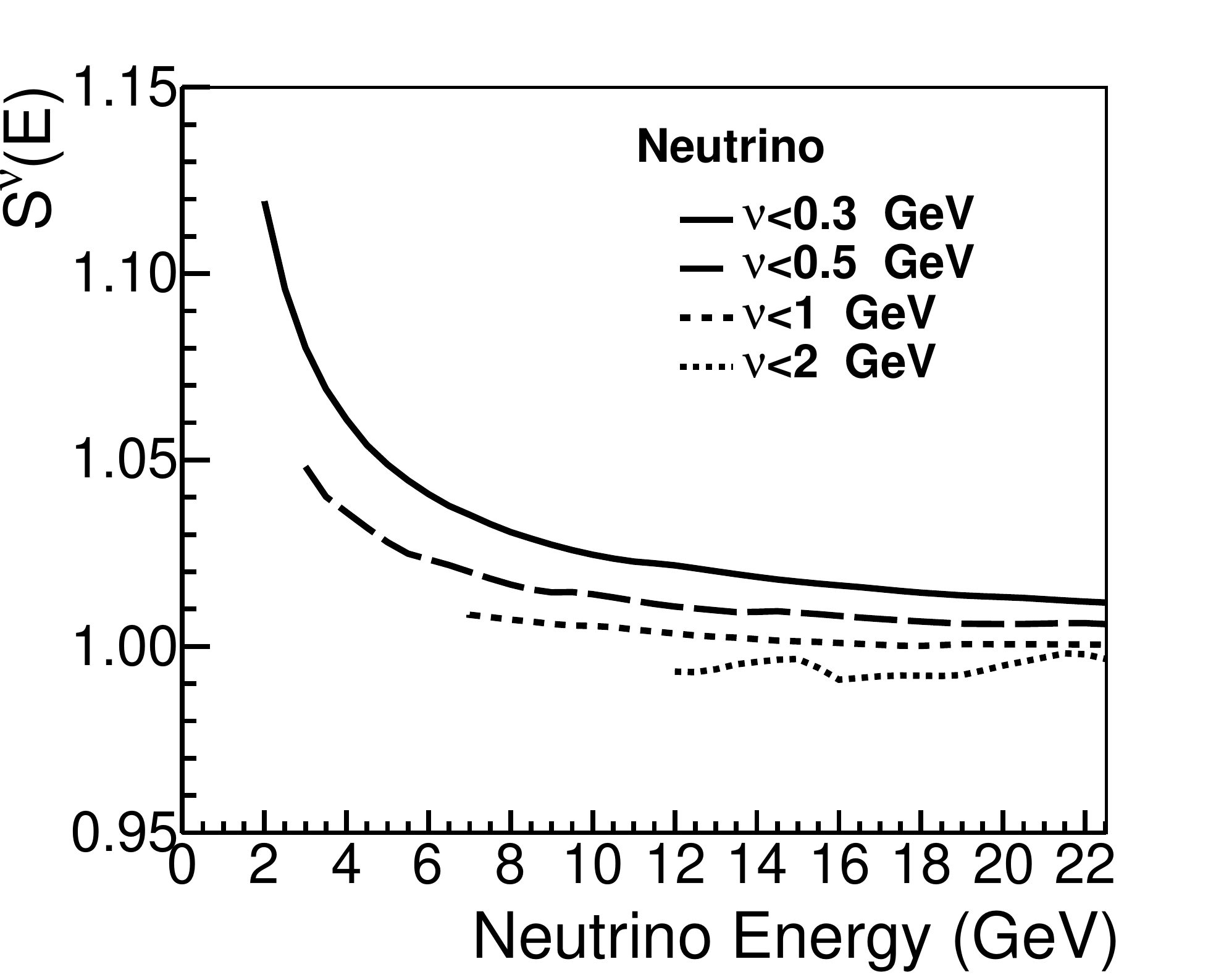}
\includegraphics[width=0.49\columnwidth]{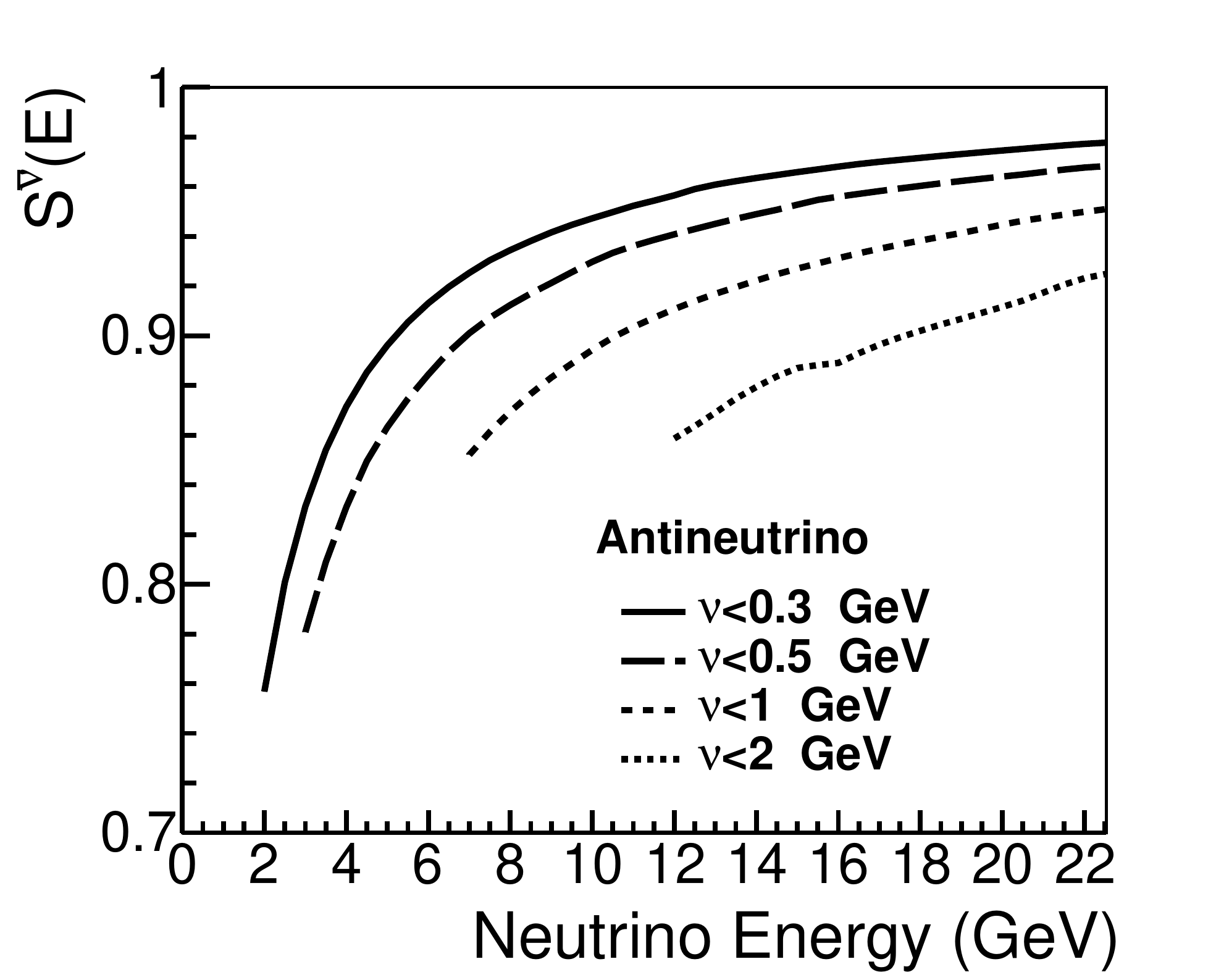}
\caption{GENIE-hybrid based low-$\nu$ corrections, \nucor ,
for neutrinos (left) and antineutrinos (right).}
\label{fig:corrections1}
\end{center}
\end{figure*}

We obtain the shape of the flux with energy from the corrected flux 
yield using Eq.~\eqref{eq:prop_flux}. 
The low-$\nu$ correction term is computed
from Eq.~\eqref{eq:lownu} using the GENIE-Hybrid model as shown in Fig.~\ref{fig:corrections1} (also in Table~\ref{tab:corrections_GENIE_nu}).

The neutrino flux is normalized using external neutrino cross section data
overlapping our sample in the normalization bin, $E_N$, (neutrino 
energies 12-22~GeV). The NOMAD~\cite{Wu:2007ab} measurement is singled
out because it is the only independent result on the same 
nuclear target (carbon) in this range. The weighted average 
value of the NOMAD from 12-22~GeV is
$\sigma_{\rm N}^{\nu}/E_{\rm N}= (0.699\pm0.025) \times 10^{-38}$cm$^{2}$/GeV.
We compute a weighted average value for our measured
unnormalized neutrino cross section, $\sigma^{\nu,\nu_0}(E_{\rm N})$,
from our points ($E=$ 13.5, 16.5, and 20~GeV) in the normalization bin
from Eq.~\eqref{eq:prop_sigma_nu}.
We obtain a normalization constant for each $\nu_0$ sub-sample, \realhnu{}, using 
\begin{equation}
\realhnu{}=\frac{\sigma^{\nu,\nu_0}(E_N) \times I^{\nu}_{iso}( E_N)}{\sigma^{\nu}_{\rm N}},
\label{eq:hnu}
\end{equation}
where the isoscalar correction, $I_{iso}$, accounts for the proton excess
($f_{p}=54\%$, $f_n=1-f_p$) in the \minerva\ target material obtained from
\begin{widetext}
\begin{equation}
I_{iso}^{\nu(\bar{\nu})}(E)=\left(\frac{\sigma_{p}^{\nu(\bar{\nu})}(E)+ \sigma_{n}^{\nu(\bar{\nu})}(E)}{f_{p} \sigma_{p}^{\nu(\bar{\nu})}(E)+f_{n} \sigma_{n}^{\nu(\bar{\nu})}(E)}\right) \left(\frac{f_{p} \sigma_{p}^{\nu(\bar{\nu})}(\nu_0,E)+f_{n} \sigma_{n}^{\nu(\bar{\nu})}(\nu_0,E)}{\sigma_{p}^{\nu(\bar{\nu})}(\nu_0,E)+ \sigma_{n}^{\nu(\bar{\nu})}(\nu_0,E)}\right).
\label{eq:iiso}
\end{equation}
\end{widetext}
Here, $\sigma_{p(n)}^{\nu(\bar{\nu})}(E)$ is the neutrino (antineutrino) cross section on a proton (neutron) in carbon and $\sigma_{p(n)}^{\nu(\bar{\nu})}(\nu_{0},E)$ is its value for $\nu<\nu_{0}$.
This correction, (see Table~\ref{tab:corrections_GENIE_nu}),
is negligible above 6~GeV and increases up to 4.2\% in the lowest energy bin.
\begin{table*}[ht]
\begin{centering}
\begin{tabular}{|c|c|c c c|c|c c c|c|c|}
\hline\hline
$\nunot$(GeV) & $E$(GeV) & $\fsamnu{}$& $\bfluxnu{}$  & $A_{\phi}^{\nu}(E)$& $H^{\nu}(\nunot)$&$\fsamnubar{}$  & $\bfluxnubar{}$& $A_{\phi}^{\bar{\nu}}(E)$& $\alpha(\nunot)$ \\
\hline
&13.50&1315 &10 & 1.18& &247 &1 & 1.04&\\
0.3&16.50&863 &4 & 1.12& 3.83$\pm$0.091&147 &1 & 0.94&1.126$\pm$0.067\\
&20.00&662 &4 & 1.05& &110 &1& 0.96&\\\hline
&13.50&2415 &15 & 1.28& &385 &2 & 1.21&\\
0.5&16.50&1613 &7 & 1.19&1.96$\pm$0.035 &224 &1 & 1.09&1.056$\pm$0.051\\
&20.00&1190 &4 & 1.16& &159 &2 & 1.12&\\\hline
&13.50&4419 &25 & 1.36& &636 &5 & 1.33&\\
1.0&16.50&2967 &12 & 1.25&1.02$\pm$0.014 &373 &3 & 1.18&1.005$\pm$0.039\\
&20.00&2235 &8 & 1.21& &260 &3 & 1.19&\\\hline
&13.50&7438 &39 & 1.42& &1033 &9 & 1.42&\\
2.0&16.50&5041 &17 & 1.28&0.574$\pm$0.006 &595 &6 & 1.23&1\\
&20.00&3826& 14 & 1.25& &427 &5 & 1.23&\\
\hline\hline
\end{tabular}
\caption{Neutrino and antineutrino flux data and corrections 
needed to apply the normalization technique described in the text.  The flux sample yield,  $F^{\nu(\bar{\nu})}$, along with 
corresponding background contribution, $B^{\nu(\bar{\nu})}_{\Phi}$, and acceptance correction, $A^{\nu(\bar{\nu})}_{\phi}$, are $\nunot$ dependent and are used to compute the unnormalized cross section.}
\label{tab:rawdata_nu_nubar_12_22}
\end{centering}
\end{table*}
\begin{table*}[ht]
\begin{centering}
\begin{tabular}{|c|c c c|c c c|}
\hline\hline
E(GeV) &$A^{\nu,KIN}_{CC}(E)$ & $S^{\nu}(\nu_{0}, E)$ &  $I^{\nu}_{iso}(\nu_{0}, E)$&$A^{\bar{\nu},KIN}_{CC}(E)$ & $S^{\bar{\nu}}(\nu_{0}, E)$ &  $I^{\bar{\nu}}_{iso}(\nu_{0}, E)$  \\
\hline
2.5 & 3.094&  1.096  &0.954 & 1.883&  0.801 &1.042\\  \hline
3.5 & 1.981&  1.040  &0.982 & 1.293&  0.809  &1.016\\
4.5 & 1.746&  1.032  &0.983 & 1.185&  0.850  &1.016\\
6 & 1.559&  1.023  &0.984 & 1.118&  0.884  &1.016\\  \hline
8 & 1.423&  1.007  &0.998& 1.076&  0.869  &1.005\\
10.5 & 1.326&  1.005  &0.998 &1.060&  0.899  &1.005\\ \hline
13.5 & 1.253&  0.995  &0.999 & 1.044&  0.875  &1.004\\  
16.5 & 1.207&  0.992  &0.999&1.035&  0.893  &1.004\\
20 & 1.171&  0.995  &0.999&1.032&  0.912  &1.004\\
\hline\hline
\end{tabular}
\caption{Neutrino and antineutrino cross section model dependent corrections computed using the GENIE-Hybrid model. $S^{\nu(\bar{\nu})}(\nu_{0}, E)$ is defined in Eq.~\eqref{eq:lownu} and $I^{\nu(\bar{\nu})}_{iso}(\nu_{0}, E)$ is defined in Eq.~\eqref{eq:iiso}.}
\label{tab:corrections_GENIE_nu}
\end{centering}
\end{table*}

In the low-$\nu$ flux extraction method, 
neutrino and antineutrino cross sections in the low inelasticity limit $y\to 0$ 
are related, and approach the same constant
value (Eq.~\eqref{eq:dsigmadnu}) for an isoscalar target in the absence of quark mixing. We make use of this 
to link the normalization of our low-$\nu$ antineutrino flux sample to that for neutrinos and therefore 
do not require external antineutrino cross section values.
The weighted average (isoscalar corrected) unnormalized antineutrino cross section, $\sigma^{\bar{\nu},\nu_0}(E_N)\times I^{\bar{\nu}}_{iso}( E_N)$, 
is computed in the normalization bin for each $\nunot$ value.
It is linked to that for neutrinos by applying a small correction due to quark mixing, which is
computed from a generator model
\begin{equation}
G(\nu_0)  =  \frac{\sigma^{\bar{\nu}}(\nu_0, E\to\infty)}{\sigma^{\nu}(\nu_0, E\to\infty)}.
\label{eq:gcorr}
\end{equation}
This correction, which is dominated by a term that is proportional to $V^2_{us}\approx 0.05$, 
is negligible for $\nu_0<0.5$~GeV, 1.5\% for $\nu_0<1$~GeV and 2.6\% for  $\nu_0<2$~GeV.
We obtain a normalization factor for the $\nu_0=2$~GeV
sub-sample from the corrected neutrino normalization, $H^{\bar{\nu}}=H^{\nu}/G$.
Rather than treating each low-$\nu$ sub-sample independently, we take 
the $\nu_0=2$~GeV value as a standard and relatively normalize among different flux samples to make them match the same value
in the normalization bin. We obtain the normalization for each $\nu_0$ sample 
from $\realhnubar{}=\realhnu{}/G(\nu_{0})/\alpha(\nu_{0})$, 
where $\alpha(\nu_0)$ is the factor needed to adjust the measured antineutrino cross section at $E_N$ to our measured value for $\nu_0=2$~GeV.
This technique makes use of additional information in our low-$\nu$ data to compensate for unmodeled cross section contributions or energy dependent systematic uncertainties in that region. 
The values of $\alpha$ (given in Table~\ref{tab:rawdata_nu_nubar_12_22}) range from 1.0 to 1.126. The size of the correction in the lowest energy bin is comparable to the size of the 1$\sigma$ systematic error in the bin (9\%). The additional statistical error from $\alpha$ is included in the result and it dominates the statistical error in the antineutrino flux and \ratio{} below 7~GeV. 

\section{Systematic Uncertainties}

We consider systematic uncertainties that arise from many sources including
muon and hadron energy scales, reconstruction-related effects, cross section modeling,
backgrounds, and normalization uncertainties.
In each case, we evaluate the effect by propagating it 
through all the steps of the analysis, including a recalculation of the absolute normalization.
The normalization technique makes the results insensitive to effects that change the overall rates.

The muon energy scale uncertainty is 
evaluated by adding the 2\% range uncertainty~\cite{Michael:2008bc}  in quadrature 
with the uncertainty in momentum measured from curvature
(2.5\% for $P_{\mu}<1$~GeV and 0.6\% for $P_{\mu}>1$~GeV), which is dominated by 
knowledge of the MINOS ND magnetic field~\cite{minervanim}.
A small component of energy loss uncertainty in \minerva\ is also taken into account. 
The hadronic response uncertainty is studied by 
incorporating an individual response uncertainty for each final state particle produced 
at the hadronic vertex in the neutrino interaction. 
A small-scale functionally-equivalent detector in a test beam~\cite{Aliaga:2015aqe} was used to assess energy responses and their uncertainties, which are found to be 3.5\% for protons, and 5\% for  $\pi^{\pm}$ and $K$.
In addition to the test beam study, information from {\it in situ} Michel electron and $\pi^0$ samples is used to 
determine the 3\% uncertainty in electromagnetic response. 
Low-energy neutrons have the largest uncertainties (25\%
for kinetic energies $<$~50~MeV and 10-20\% for $>$~50~MeV), which are estimated by benchmarking GEANT4{~\cite{Agostinelli:2002hh} 
neutron cross sections against $nA\rightarrow pX$ measurements in this energy range. 
The energy scale uncertainties are the most important components of the flux shape measurement, but these largely cancel in cross sections and \ratio{}, resulting in a smaller overall effect.

Two reconstruction-related sources of uncertainty that affect measured shower energies were considered. The effect of PMT channel cross-talk is studied  by injecting cross-talk noise into the simulation and its uncertainty is estimated by varying the amount by 20\%. The resulting uncertainty is small and is added in quadrature with the hadronic energy scale uncertainty.  Muon track-related energy depositions (from $\delta$-rays or bremsstrahlung) are difficult to isolate within the shower region. We use data and simulation samples of beam-associated muons passing through the detector to model these and tune our hadron energy distribution in data and simulation. We compare two algorithms to separate muon-associated energy from the shower region and take their difference as the uncertainty from this source, which is also found to be small.

The effect of accidental activity from beam-associated  muons is simulated by overlaying events from data within our reconstruction timing windows. 
We study overall reconstruction efficiency as a function of neutrino energy by projecting track segments reconstructed using only the \minerva\ detector and searching for the track in MINOS ND, and vice versa.
Track reconstruction efficiency, which agrees well between data and Monte Carlo, is above 99.5\%  for \minerva\ and above 96\%
for MINOS ND and is 
found to be nearly constant
with energy. We adjust the simulated efficiency accordingly, although the
normalization procedure makes the results insensitive to these effects.

Cross section model uncertainties enter into the measurement directly through the model-dependent correction as well as through bin migration effects at the boundaries of our experimental acceptance. Our default model (GENIE-Hybrid) is based on GENIE 2.8.4, we therefore use the prescription in Ref.~\cite{Andreopoulos:2015wxa} to evaluate uncertainties on all of the corresponding model parameters. The largest GENIE model uncertainties arise from final state interactions (FSI) and the resonance model parameters. We account for uncertainties in the resonance contribution by varying the axial mass parameters, $M^{RES}_A$ and $M^{RES}_V$, in our model by $\pm 20\%$ and $\pm 10\%$, respectively. The resulting effect on the cross section is up to $4\%$. 
The GENIE parameters that control FSI effects include mean free path, 
reaction probabilities, nuclear size, formation time and hadronization model variation.
The largest FSI uncertainty, due to the pion mean free path within the nucleus,  is up to 2\% (3\%) for cross sections (fluxes).
We separately evaluate the uncertainties from the tuned model components (RPA, single pion non-resonant, and 2p2h) discussed
in Sec.~\ref{sec:xsec}. We include half the difference between the default GENIE 2.8.4 and the implemented RPA model in quadrature into the total model uncertainty. We assume a 15\% uncertainty in the retuned non-resonant single pion production component. 
After incorporating the 2p2h model, a sizable discrepancy in the hadronic energy distribution with the data remains. To assess an additional uncertainty from this unmodeled contribution, we fit the data excess at low hadronic energy described in Ref.~\cite{Rodrigues:2015hik} in the neutrino energy range $2<E_{\nu}<6$~GeV (taking into account separately proton-proton and proton-neutron initial states) to obtain a corrected model~\cite{luren,ddref}. 
We take the  uncertainty as the difference of the result obtained with this data-driven model, from the nominal result. The \minerva\ antineutrino data also show an excess in the same region. We apply the corrected model from neutrino described above and then fit the remaining antineutrino excess to obtain a data-driven antineutrino 2p2h model uncertainty. The primary effect of varying the size of this contribution is to shift  the overall level of the cross section. The normalization procedure removes most of the effect and the remaining uncertainty is less than 1.5\% (2\%) on the cross section (flux).

The contamination from wrong-sign events is significant only for
the antineutrino sample (about 4\% above 15 GeV). To evaluate the uncertainty from this source we recompute the antineutrino cross section with wrong-sign 
events in RHC mode reweighted by the extracted neutrino \lownu{} flux. The difference is taken as the wrong-sign contamination uncertainty, 
which is less than 0.5\% ($0.2\%$) for the extracted antineutrino cross section (flux).

The overall 3.6\% normalization uncertainty arises from the precision of the NOMAD data set in the energy range 12-22~GeV. We have assumed
NOMAD data points in this region to have 100\% correlated point-to-point systematic uncertainties in computing the weighted average error from their data.
For antineutrinos and \ratio{} we study an additional contribution to the uncertainty from the
correction term, $G(\nunot)$, by varying the GENIE-Hybrid cross section model parameters within their uncertainties prescribed by GENIE. 
The resulting uncertainty is negligible (less than 0.5\% for all energies). 

An error summary for the fluxes is shown in Fig.~\ref{fig:error_flux}. The dominant systematic uncertainties 
on the shape for both the neutrino and antineutrino fluxes arise from limited knowledge of muon and hadron energy scales. This uncertainty
peaks at low energies and has a nontrivial energy dependence
that is due to the combined effects from sub-components having different precisions, as well as to the flux shape itself. The FSI uncertainty gives an effect that is also important, 3.5\%, and nearly constant with energy. 
For antineutrinos, the statistical precision is poorer and is comparable to the systematic precision over most of the energy range.
The statistical error in the
data-based cross normalization factor $\alpha(\nunot)$ (Table~\ref{tab:rawdata_nu_nubar_12_22}), dominates the statistical precision below 12~GeV and is responsible
for the detailed shape features in the uncertainty band\footnote{Features occur where the $\nunot$ cut value changes at 3, 7, and 12~GeV.}.

\begin{figure*}[!htb]
\begin{center}
\includegraphics[width=0.49\columnwidth]{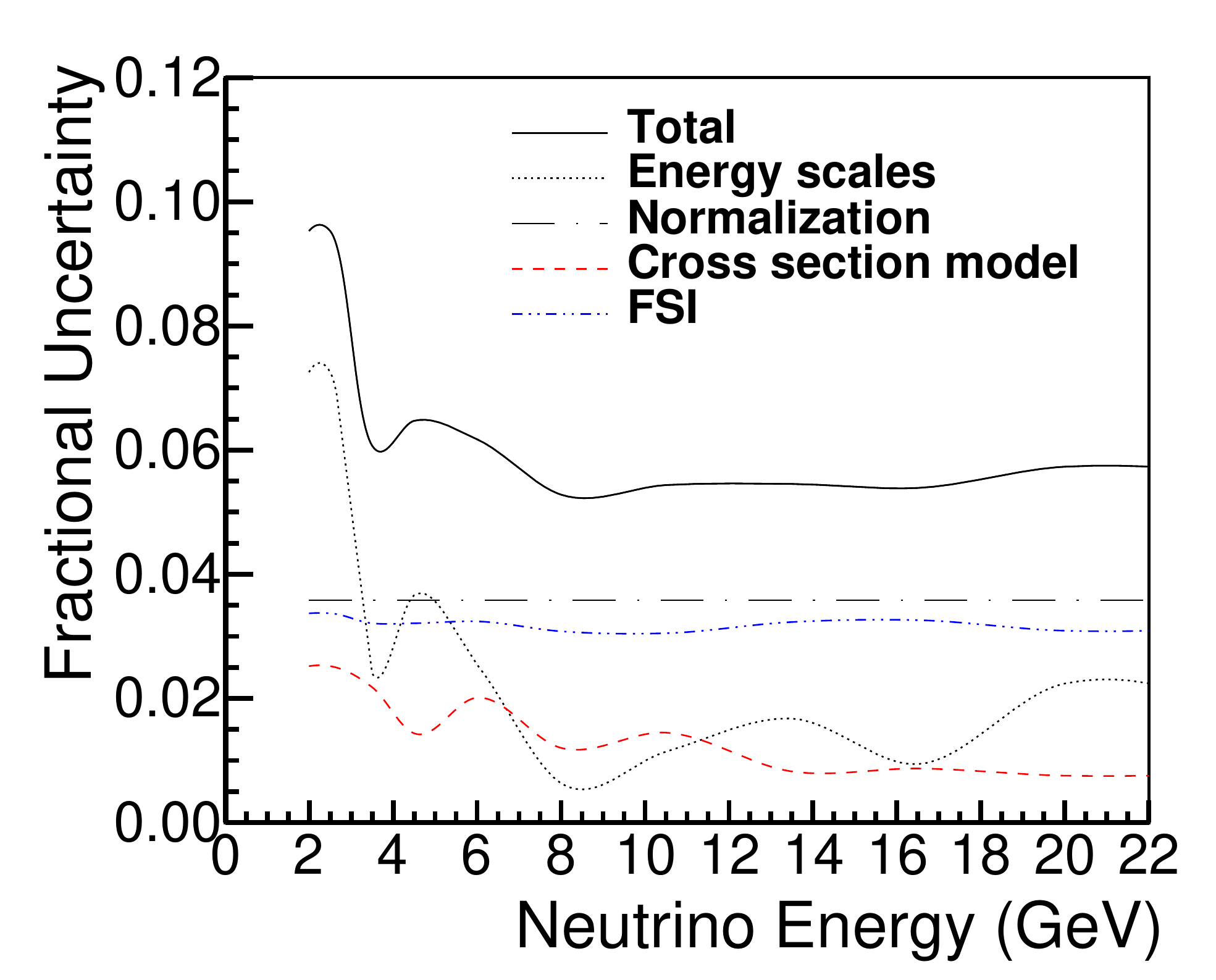}
\includegraphics[width=0.49\columnwidth]{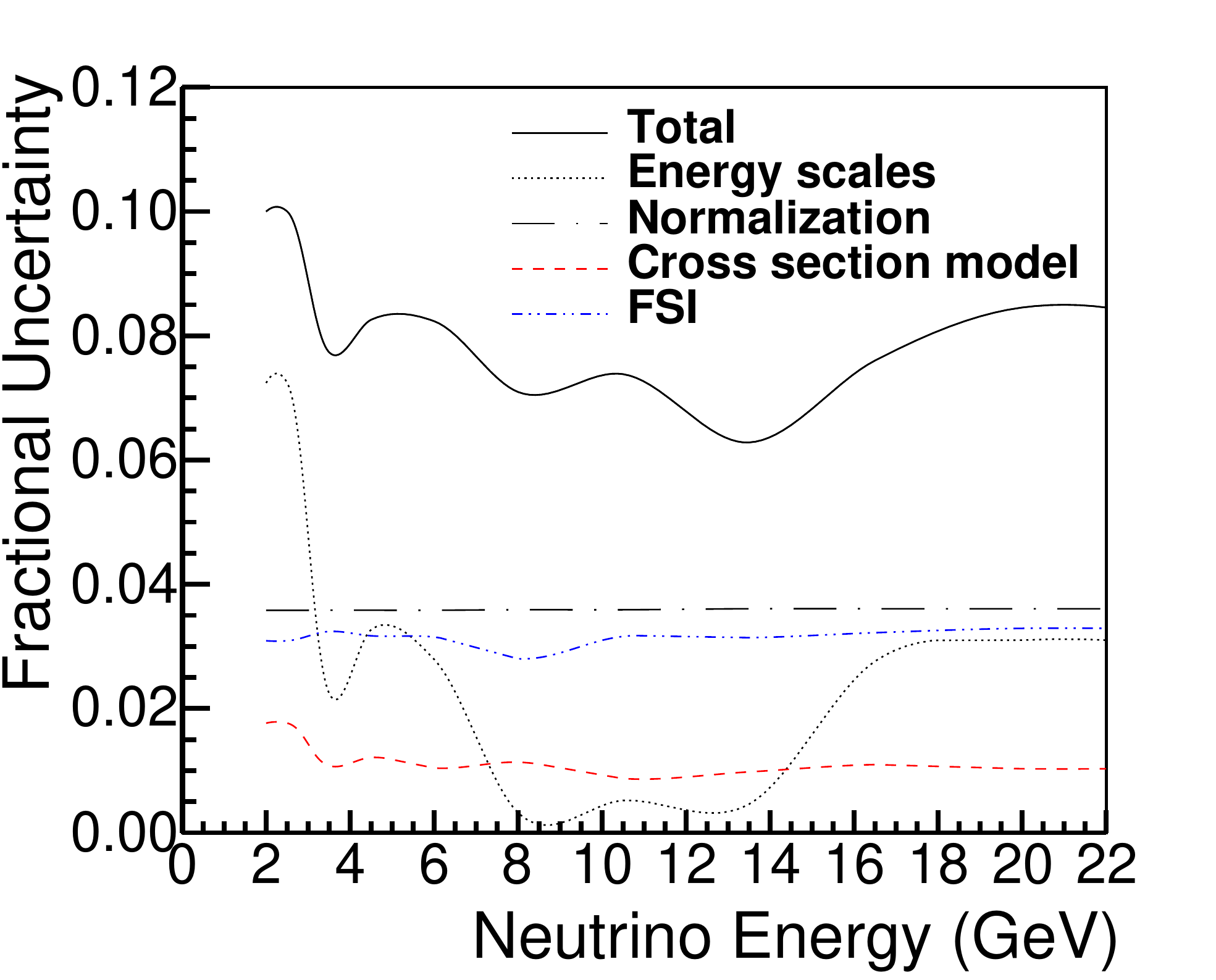}
\caption{Measurement uncertainties for neutrino (left) and antineutrino (right) low-$\nu$ fluxes. 
The total uncertainty (sys.~+~stat.) is the solid line. Components 
from cross section model (dashed red), FSI (dot-dash blue), and energy scales (dotted) are shown. The 3.6\%  uncertainty in the external normalization (dashed black) is the error of the NOMAD data in the normalization region. }
\label{fig:error_flux}
\end{center}
\end{figure*}

Neutrino and antineutrino cross section uncertainty components are summarized in Fig.~\ref{fig:error_xsec}.
Many systematic effects cause changes that are similar in the
cross section and flux samples and partially cancel
in the measured cross section. The dominant uncertainty is from the cross section model
at low energy, while normalization dominates at high energies. 
Neutrino and antineutrino cross sections have comparable systematic errors 
but the statistical precision is poorer for antineutrinos and 
it dominates the error in all but the lowest energy bin.

\begin{figure*}[!htb]
\begin{center}
\includegraphics[width=0.49\columnwidth]{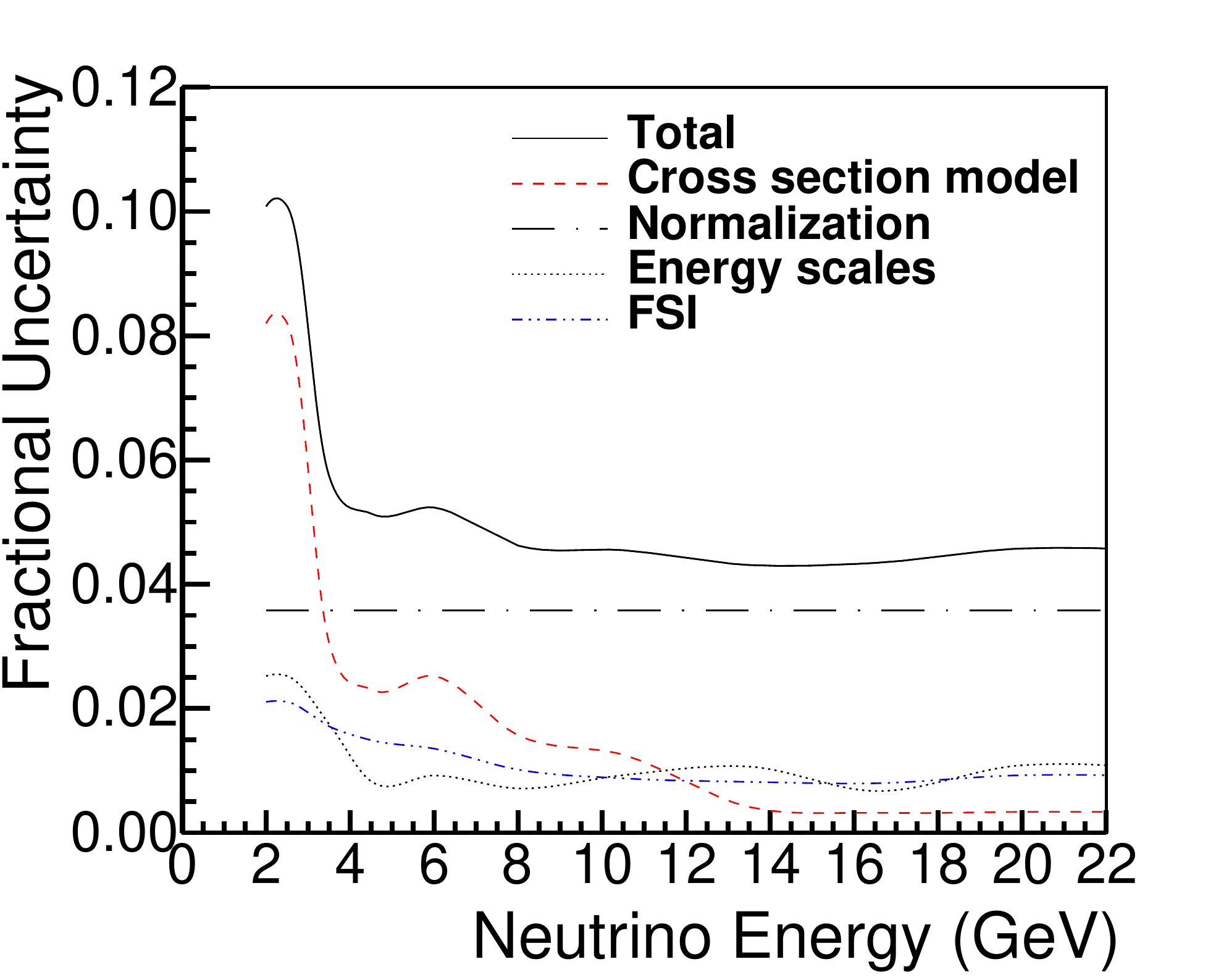}
\includegraphics[width=0.49\columnwidth]{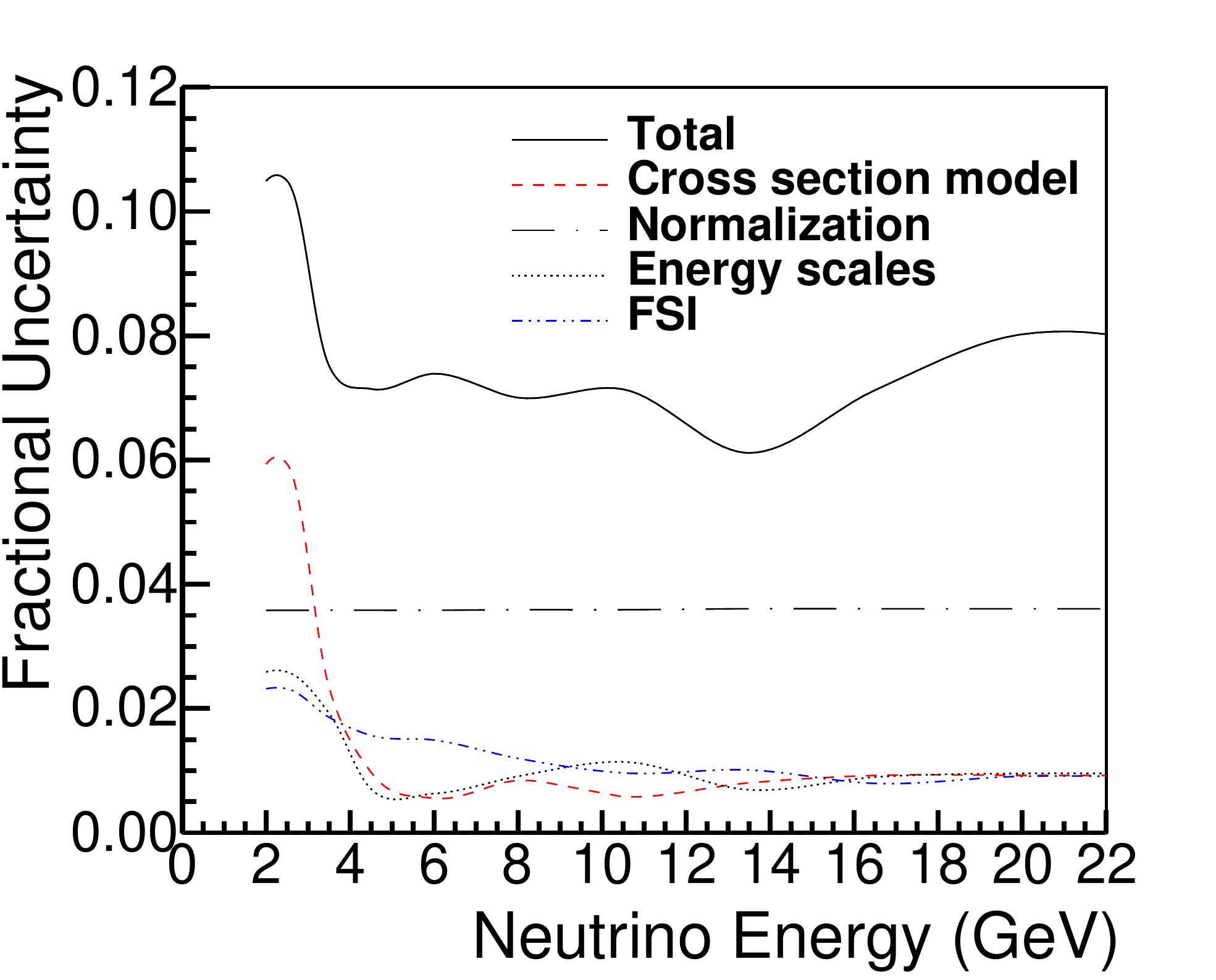}
\caption{Measurement uncertainties for neutrino (left) and antineutrino (right) total cross sections. 
The total uncertainty (sys.~+~stat.) is the solid line. Components 
from the cross-section model (dashed red), FSI (dot-dash blue), and energy scales (dotted) are shown. The 3.6\%  uncertainty in the external normalization (dashed black) is the error of the NOMAD data in the normalization region.
Statistical error dominates the measurement in the antineutrino result.}
\label{fig:error_xsec}
\end{center}
\end{figure*}
The uncertainties on the cross section ratio, \ratio{}, are summarized in Fig.~\ref{fig:error_r}. Energy scale uncertainties nearly cancel 
in this ratio, and the sizes of effects from FSI and many model uncertainties are reduced.
The dominant remaining uncertainties are from the 
$M^{RES}_A$ cross section model parameter 
and the effect of implementing the RPA model in GENIE 2.8.4.  
The corresponding cross section components produce sizable shape effects in the visible energy in the low-$\nu$ region.
Different final states in neutrino versus antineutrino interactions reduce cancellation effects 
in these components for the ratio.
The overall uncertainty in \ratio{} is dominated by statistical uncertainty in the antineutrino sample.
\begin{figure}[!htb]
\begin{center}
\includegraphics[width=0.49\columnwidth]{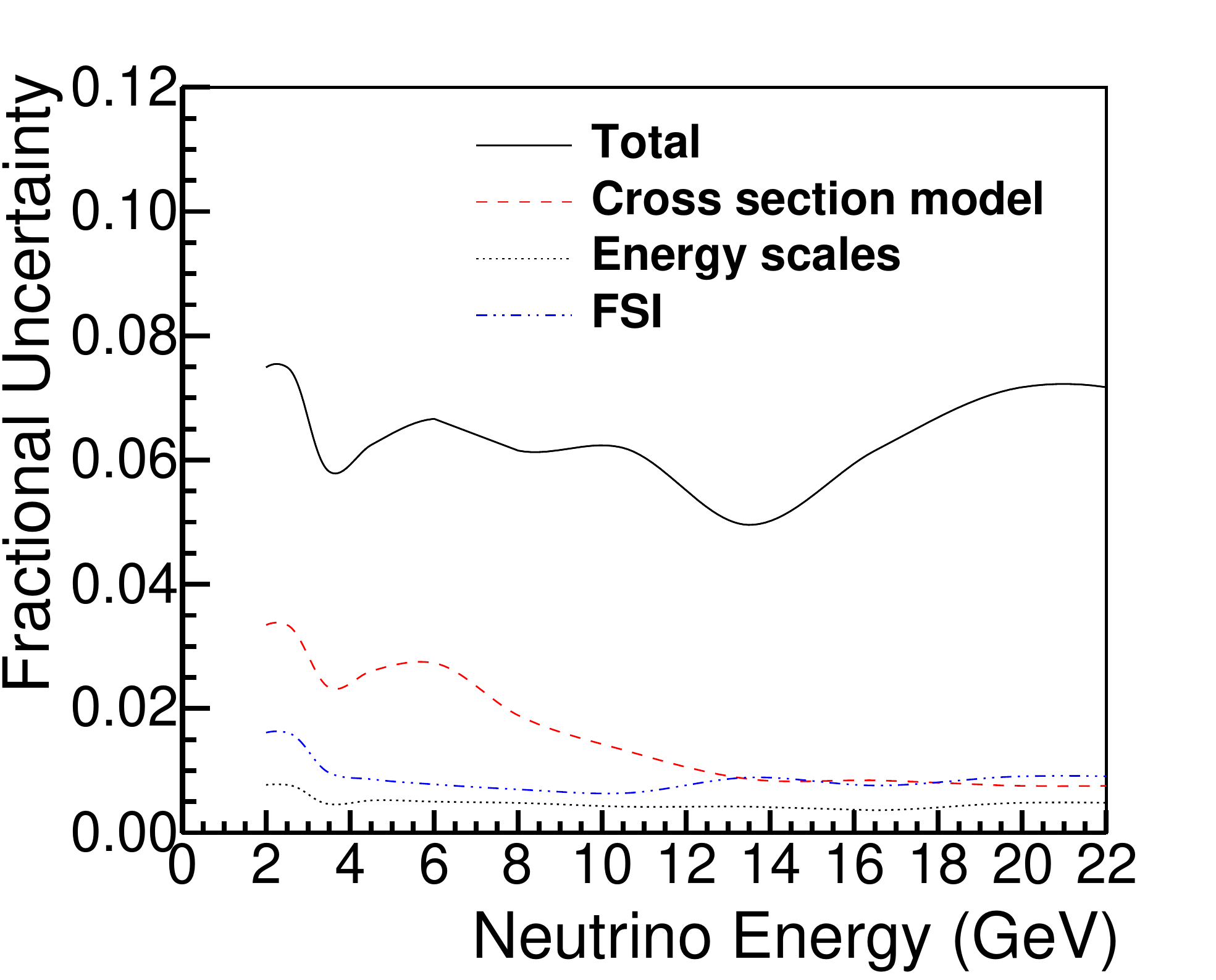}
\caption{Measurement uncertainties for the cross section ratio, \ratio{}.
The total uncertainty (sys.~+~stat.) is the solid line. Components 
from cross section model (dashed red), FSI (dot-dash blue), and energy scales (dotted) are shown. Normalization uncertainty 
is very small ($<$1\%) and is included in the total error curve. 
The uncertainty is dominated by the statistical precision of the antineutrino sample.}
\label{fig:error_r}
\end{center}
\end{figure}

\section{Flux and Cross Section Results}

The extracted \lownu{} flux (Table~\ref{tab:final_results}) is shown in Fig.~\ref{fig:fluxes} where it is compared to the \minerva\ simulated flux of Ref.~\cite{Aliaga:2016oaz}. The latter flux is constrained using hadron production data  and a detailed GEANT4{~\cite{Agostinelli:2002hh} beamline simulation. The extracted flux low-$\nu$ is in reasonable agreement with the simulation for both modes\footnote{Our previous measurement uses an earlier version of the simulated flux as described in~\cite{DeVan:2016rkm}.}. The low-$\nu$ measurement prefers a smaller neutrino flux below 7~GeV (approximately 5\%) while a larger flux is preferred for both neutrinos and antineutrinos (2-12\% for neutrinos, up to 16\% for antineutrinos) in the $>$7~GeV range.
The low-$\nu$ flux compared to the flux of the tuned production-based simulation achieves better precision for neutrinos (by 30\% for $E_{\nu}$ above 3~GeV) and comparable for antineutrinos.
\begin{figure*}[!htb]
\begin{center}
\includegraphics[width=0.49\columnwidth]{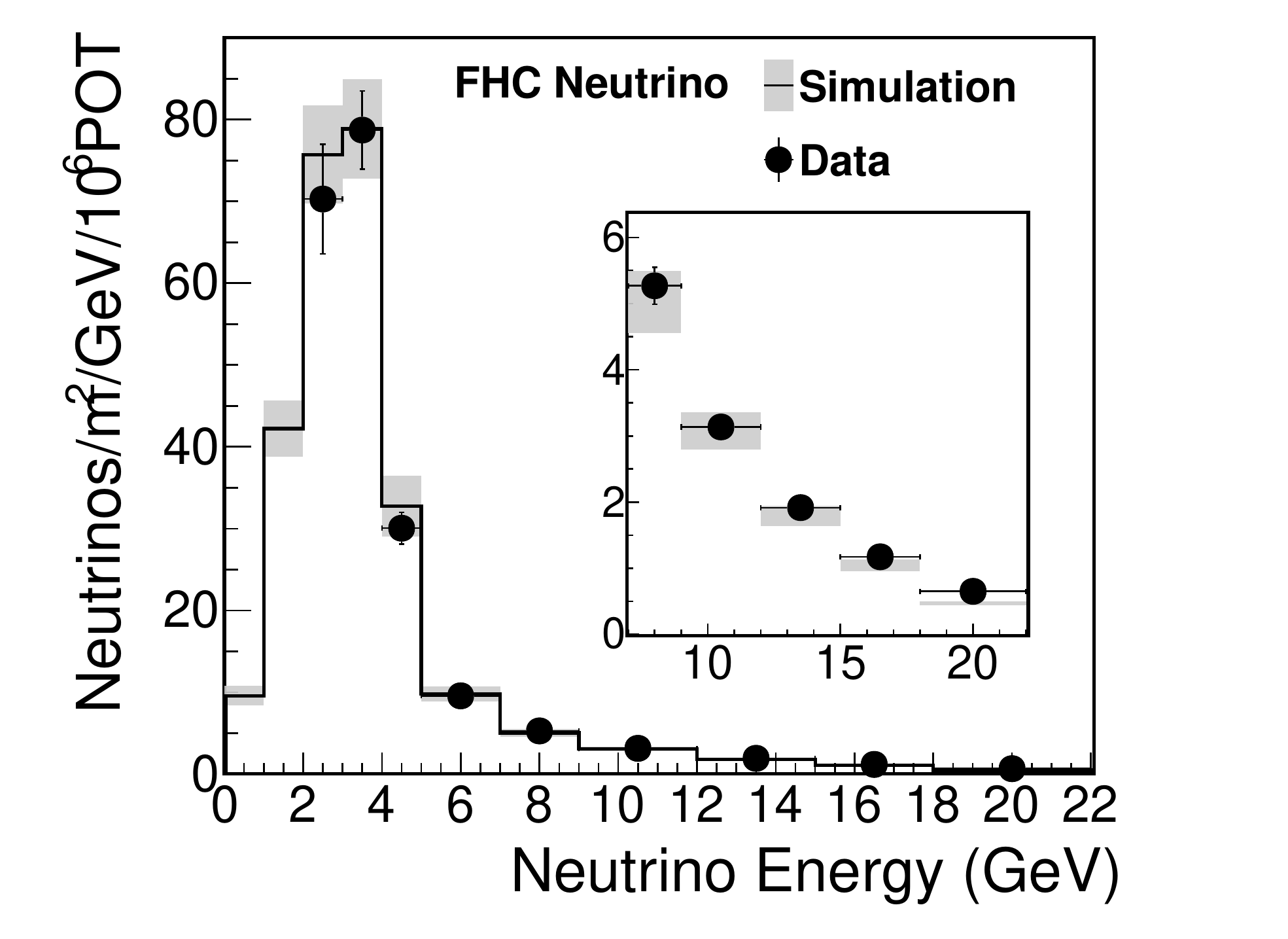}
\includegraphics[width=0.49\columnwidth]{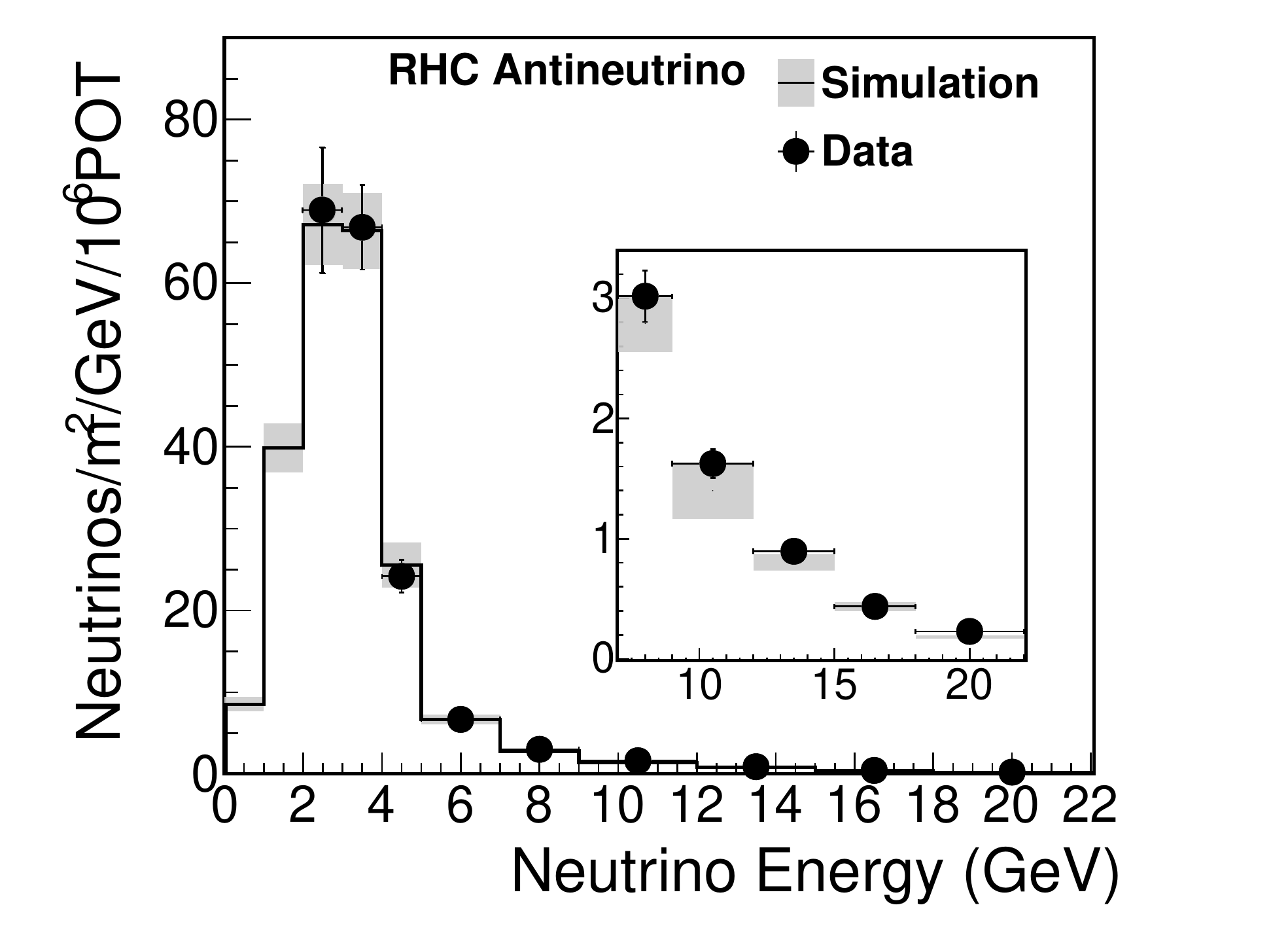}
\caption{Extracted \lownu{} flux (points) for FHC neutrino (left) and RHC antineutrino (right). The histogram shows the Monte Carlo simulated fluxes
from Ref.~\cite{Aliaga:2016oaz} and one sigma error band (shaded bars). The insets show a zoom-in of the 7-22~GeV energy range.}
\label{fig:fluxes}
\end{center}
\end{figure*}

The measured cross sections (Table~\ref{tab:final_results}) are shown in Fig.~\ref{fig:xsecs} compared with the GENIE-Hybrid model. 
\begin{figure*}[!htb]
\begin{center}
\includegraphics[width=0.49\columnwidth]{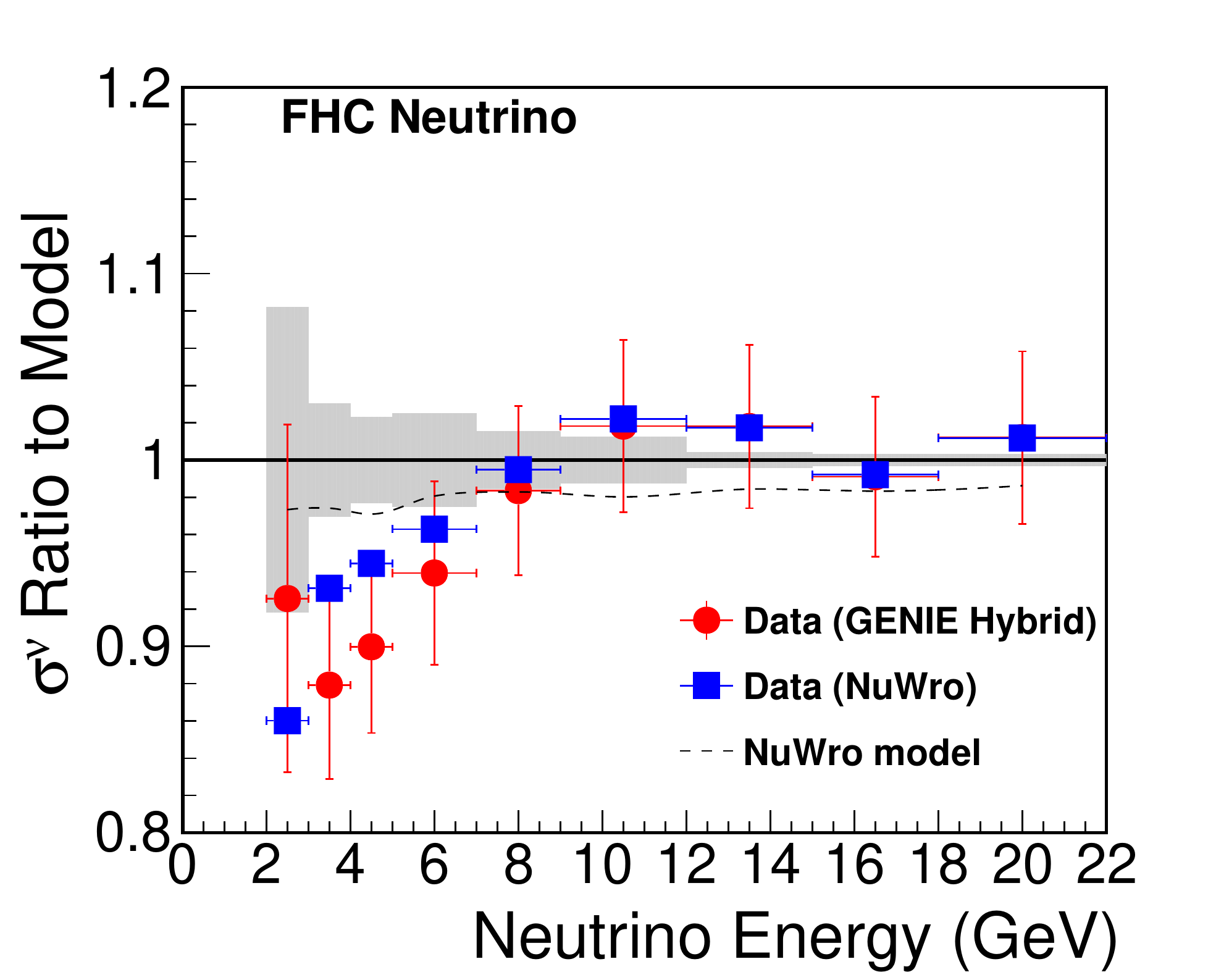}
\includegraphics[width=0.49\columnwidth]{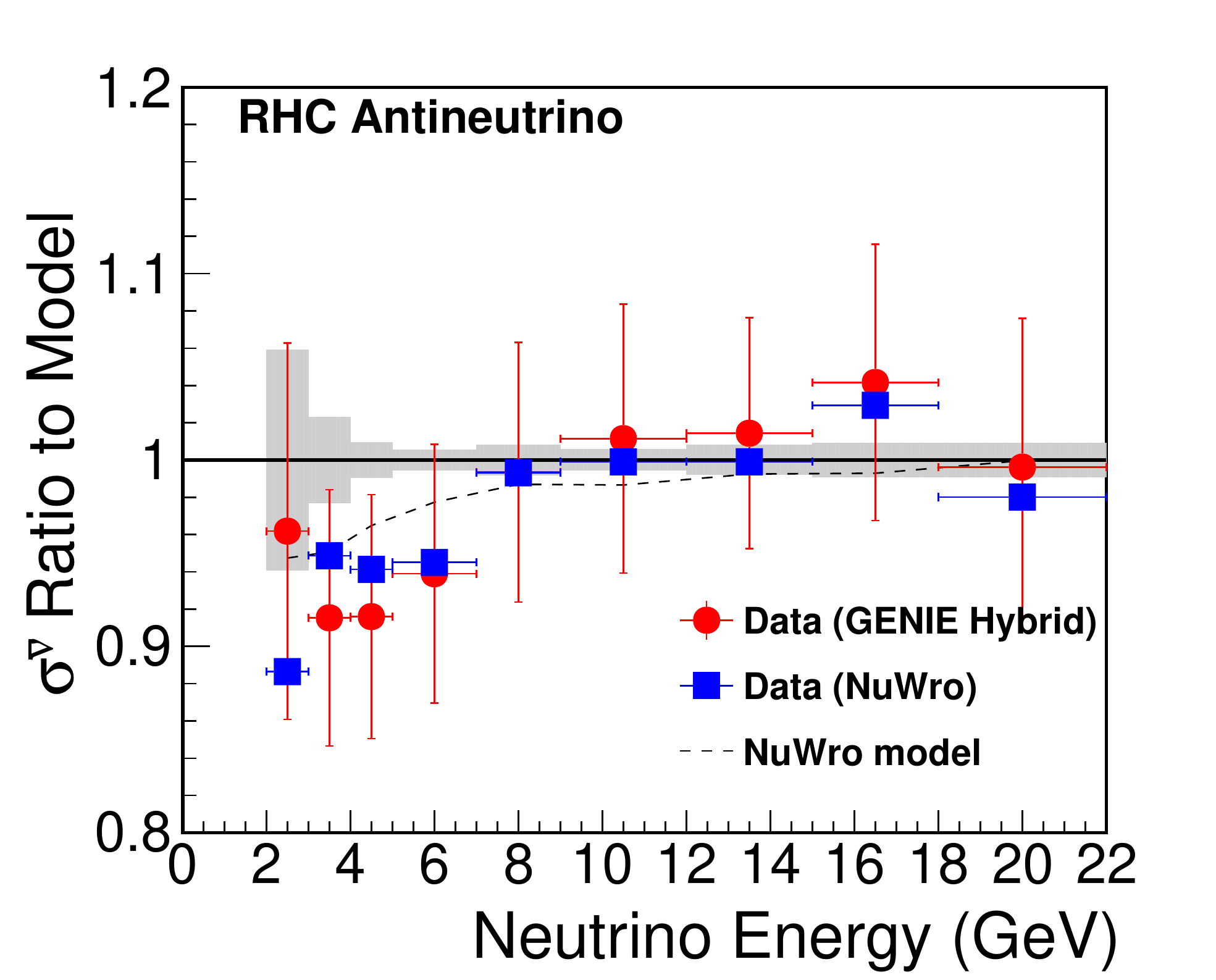}
\caption{Ratio of the measured neutrino (left) and antineutrino (right) cross sections to the GENIE-Hybrid model. 
Points are \minerva\ data with default GENIE-Hybrid (circles) and alternative NuWro model (squares) 
used to compute model-based correction terms. GENIE-Hybrid data points are plotted with total error (sys.~+~stat.). The dashed line shows the NuWro model. 
The shaded band give the size of the cross section model systematic uncertainty.}
\label{fig:xsecs}
\end{center}
\end{figure*}
The data (red points), extracted using GENIE-Hybrid for model corrections, favor a lower total cross section
in the region 2-9~GeV, where data lie below the curves (by up to $\sim$2$\sigma$) for neutrinos. Antineutrino data
also favor a lower cross section in the same region, but agree with models within the 
precison of the data, which have larger statistical uncertainties. 
For comparison, we also extract results using Eqs.~\eqref{eq:prop_sigma_nu} and \eqref{eq:ratio} and 
NuWro (squares)
to compute explicit model correction terms\footnote{
GENIE 2.8.4 with FSI turned on is used to simulate the 
fully reconstructed \minerva\ samples, and to correct for detector effects we deliberately turn the FSI processes off in NuWro, to avoid double counting them.}.
We omit error bars from NuWro-based points, which use the same raw binned data, and  therefore  have the same 
(correlated) statistical and detector-related 
systematic uncertainties. 
The shaded band shows the size of the estimated model systematic uncertainty (computed from the GENIE-Hybrid model)
which spans the differences between the extracted cross section values.
The NuWro model has a different
treatment of the low-$\nu$ region than GENIE, 
including a different axial mass parameter ($M_A=1.2$~GeV), a transverse enhancement model (TEM)~\cite{Bodek:2011ps}) to account
for the meson exchange current (MEC) scattering contribution, and a duality-based treatment in the resonance region~\cite{Graczyk:2005uv}.
The two sets of extracted cross sections show significant
differences at low energies that reflect
different modeling of the kinematic acceptance 
correction ($A^{KIN}_{CC}$), which is larger for $E_{\nu}<7$~GeV. 
QE and MEC components, which dominate the lowest energy bin, 
have a harder muon spectrum resulting in better acceptance in the NuWro model. 
GENIE kinematic acceptance is better in the 3-7~GeV energy range for the
resonance and deep inelastic scattering (DIS) components, which 
become dominant above 3~GeV. 
At high energies, the normalization method removes the effect of correction differences
between the two models for the neutrino data points. 
For antineutrinos,
the GENIE-Hybrid results are systematically above those for the NuWro model by a few percent at high energies. 
We have applied the GENIE-Hybrid quark mixing correction $G(\nunot)$
to the NuWro data points, which does not include quark mixing by default.
Figure~\ref{fig:world_xsecs} shows a comparison of the measured 
charged-current total cross sections with 
world neutrino data~\cite{minos, Eichten:1973cs, Barish:1976bj, Barish:1978pj, Baltay:1980pr, Baker:1982ty, Ciampolillo:1979wp, Baranov:1978sx, Mukhin:1979bd, Anikeev:1995dj, Wu:2007ab, Nakajima:2010fp, Anderson:2011ce, Abe:2013jth, Acciarri:2014isz, Abe:2015biq}.
We apply a non-isoscalarity correction\footnote{Corrections for SciBooNE CH target points with energies in the range 0.38-2.47~GeV are 1.085, 1.06, 1.038, 1.033, 1.028, 1.028, respectively. 
We correct T2K 2013 (CH target at  E=0.85~GeV) by 1.04, T2K 2014 (iron at E=1.5~GeV) by 0.977, T2K 2015 (iron at E=1~GeV, 2~GeV, and 3~GeV) 
by 0.971, 0.976 and 0.977, respectively.} to other data sets to compare 
with our isoscalar-corrected carbon measurement.
The neutrino cross section is in good agreement with other measurements 
that overlap in this energy range and is among the most precise in the resonance-dominated region (2-7~GeV). 
Comparisons with world antineutrino data~\cite{minos, Mukhin:1979bd, Asratian:1978rt, Fanourakis:1980si} are also shown.
Our data add information in the region below 10~GeV where previous antineutrino data are sparse
and improve precision and coverage, especially in the region below 6~GeV. Our results are in agreement with precise
data on other nuclei~\cite{minos} in the neutrino energy region of overlap ($>6$ GeV) and provide the most precise measurement of the antineutrino cross section below 5~GeV to date.
\begin{figure*}[!htb]
\begin{center}
\includegraphics[width=\columnwidth]{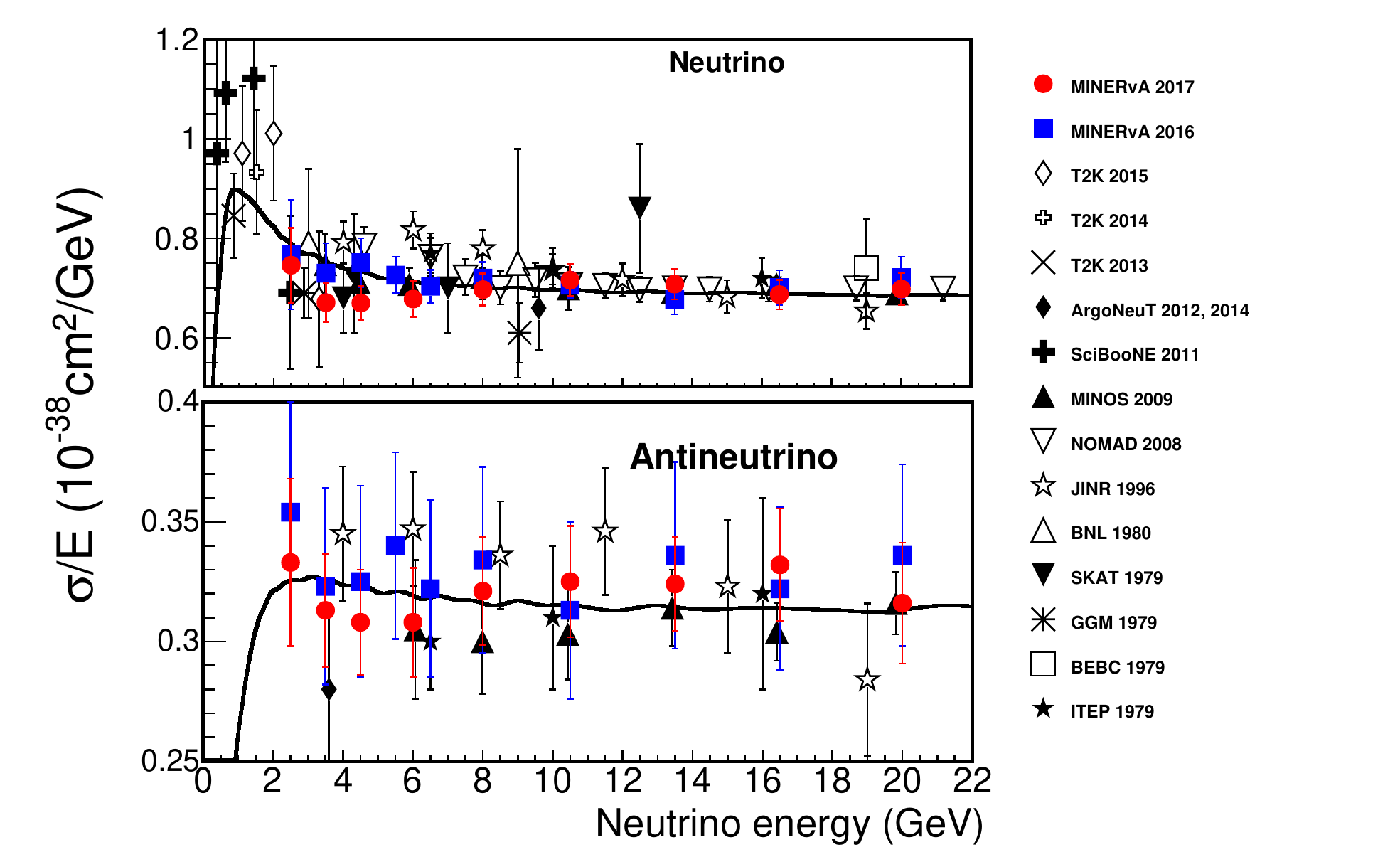}
\caption{\minerva measured neutrino and antineutrino charged-current 
inclusive cross sections (red circles and previous result from Ref.~\cite{DeVan:2016rkm}
 shown with blue squares) compared with other measurements for neutrinos~\cite{minos, Eichten:1973cs, Barish:1976bj, Barish:1978pj, Baltay:1980pr, Baker:1982ty, Ciampolillo:1979wp, Baranov:1978sx, Mukhin:1979bd, Anikeev:1995dj, Wu:2007ab, Nakajima:2010fp, Anderson:2011ce, Abe:2013jth, Acciarri:2014isz, Abe:2015biq} (upper plot), and antineutrinos ~\cite{minos, Mukhin:1979bd, Asratian:1978rt, Fanourakis:1980si} (lower plot), on various nuclei in the same energy range. The reference curve shows the prediction of GENIE 2.8.4.}
\label{fig:world_xsecs}
\end{center}
\end{figure*}

The measured cross section ratio, \ratio{}, is shown in Fig.~\ref{fig:ratio} compared with GENIE and NuWro models and with world data~\cite{minos}, \cite{Eichten:1973cs}, \cite{Mukhin:1979bd}. 
\begin{figure*}[!htb]
\begin{center}
\includegraphics[width=0.49\columnwidth]{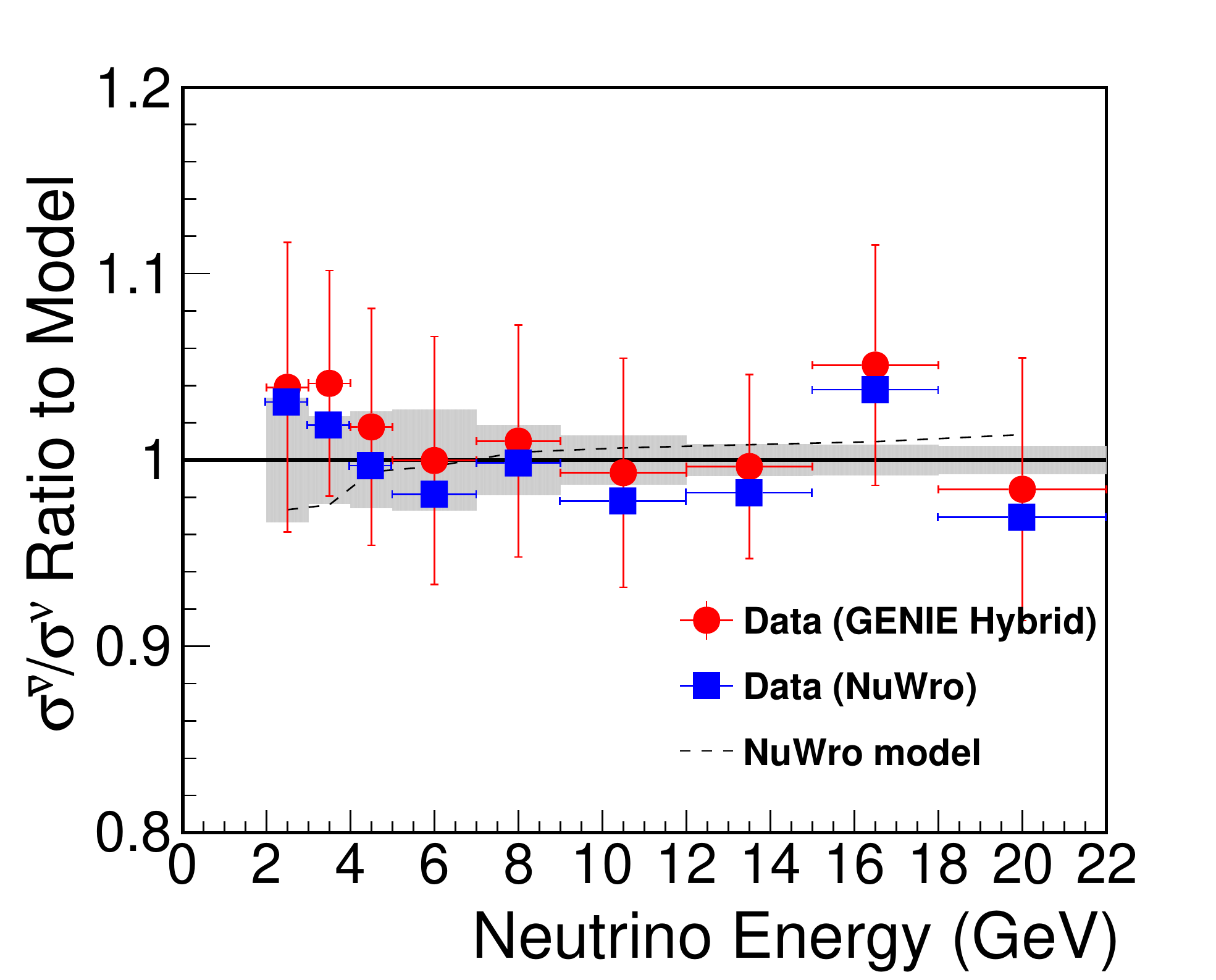}
\includegraphics[width=0.49\columnwidth]{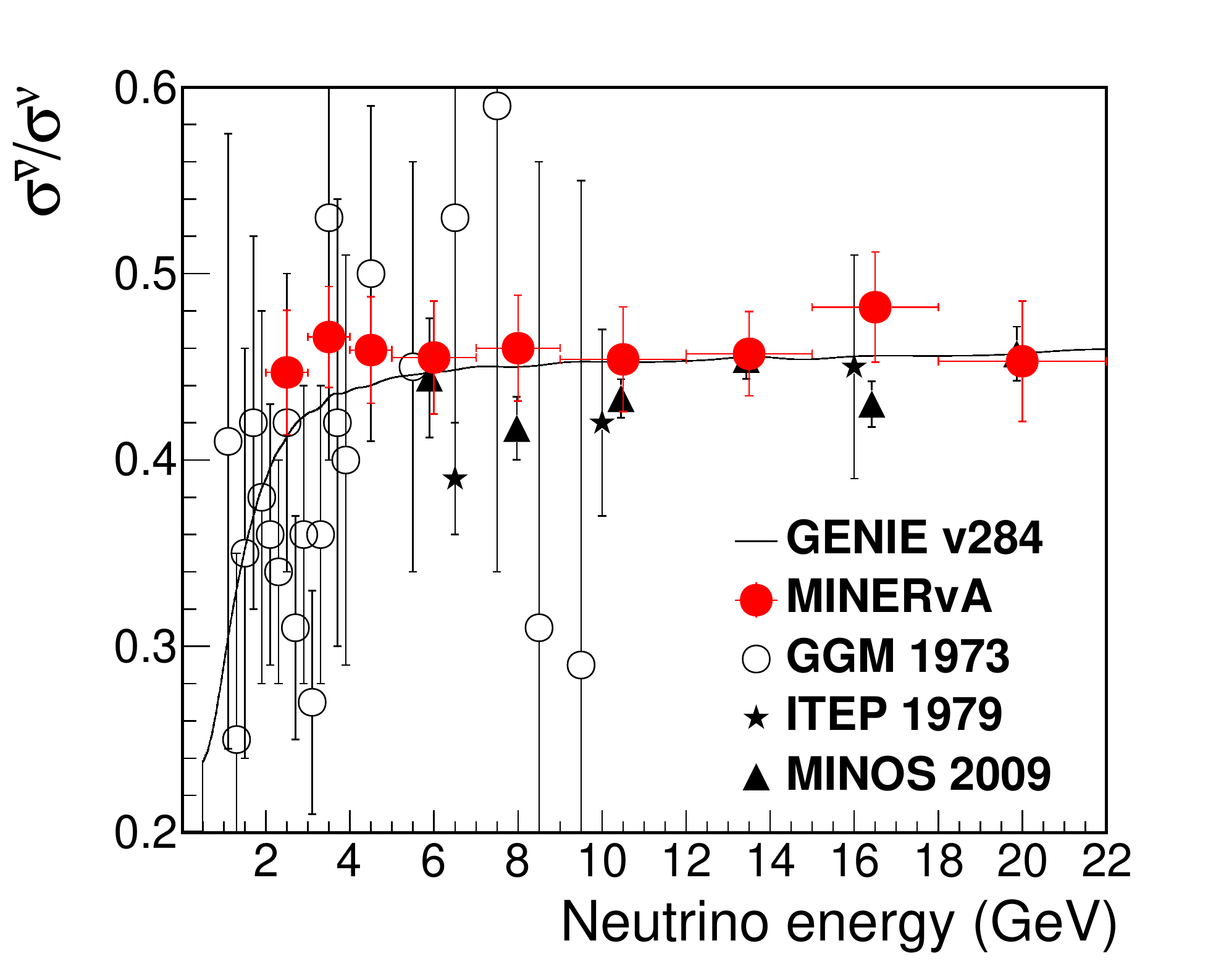}
\caption{(Left) Ratio of measured \ratio{} to GENIE-Hybrid. Points are \minerva\ data with default GENIE-Hybrid (circles) and alternative NuWro model (squares) used to compute model-based correction terms. GENIE-Hybrid data points are plotted with total error (sys.~+~stat). The dashed line shows the NuWro model. 
The shaded band shows the size of the cross section model systematic uncertainty. (Right) Comparison of \minerva\  \ratio{}
(corrected to an isoscalar target) with world measurements(~\cite{Eichten:1973cs}, \cite{Mukhin:1979bd} and \cite{minos}).
}
\label{fig:ratio}
\end{center}
\end{figure*}
Measured points are extracted using GENIE-Hybrid (circles) and NuWro (squares) for model corrections.
The measured \ratio{} lies above the model predictions at low energies and 
favors a flatter extrapolation into that region than do the models, which fall off below 5~GeV.
The NuWro results are systematically below the  GENIE-Hybrid results by a few percent, tracking 
the differences seen in the antineutrino cross section level in the numerator (discussed above).   
The differences between GENIE-Hybrid-based and NuWro-based \ratio{} measurements 
at lower energies are less significant
than differences seen in the cross sections from the two models.
The shaded band, which spans the NuWro versus GENIE-Hybrid point differences,
shows the size of the  estimated systematic uncertainty from model sources. 
Our result is in good agreement with the recent measurement from MINOS on an iron target 
in the region where they overlap ($E_{\nu} >~$6~GeV). This measurement is the only precise determination of \ratio{} in the $E_{\nu}<~$6~GeV region. It spans neutrino energies from 2 to 22~GeV, a range which is highly relevant to ongoing and future oscillation experiments.

\section{Conclusion}

We present the first precise measurement of the ratio of antineutrino to neutrino cross sections, \ratio{}, in the region below 6~GeV, which is important for future long baseline neutrino oscillation experiments. Our measurement, with precision in the range of 5.0-7.5\%, represents an improvement by nearly a factor of four over 
the previous measurements in this region~\cite{Eichten:1973cs}.
We measure neutrino and antineutrino cross sections that extend the reach for antineutrino data to low energies and are among the most precise in the few GeV energy range. Two leading neutrino generators, GENIE and NuWro, both overestimate the measured 
inclusive CC cross sections at the level of 4-10\% as energy decreases from 9~GeV to 2~GeV.
We also present measured total and low-$\nu$ fiducial rates that can be used to obtain the cross sections and their ratio with other models.
In the near future, this will allow our data to be used with new models that will have improved treatments of nuclear effects and low energy scattering processes.

The cross section ratio \ratio{} is found to have systematic uncertainties that are significantly smaller than those associated with either of the CC inclusive cross sections, due to cancellation of common systematic uncertainties. 
We demonstrate the robustness of \ratio{} by comparing results using two different models 
(GENIE-Hybrid and NuWro). The differences are found to be smaller than in the individual cross
section measurements and are comparable with the size of estimated model systematic uncertainties.

\begin{sidewaystable}
\label{tab:final_results}	
\begin{centering}
\begin{tabular}{|c||c c c c|c c c c||c c c c|c c c c||c c c c|}
\hline\hline
E & \multicolumn{4}{c|}{ $\Phi^{\nu}(E)$} & \multicolumn{4}{c||}{ $\sigma^{\nu}(E)/E$}& \multicolumn{4}{c|}{$\Phi^{\bar{\nu}}(E)$} & \multicolumn{4}{c||}{$\sigma^{\bar{\nu}}(E)/E$}&\multicolumn{4}{c|}{\ratio{}}\\
GeV & \multicolumn{4}{c|}{ $neutrinos/m^{2}/GeV/10^{6}POT$} & \multicolumn{4}{c||}{$10^{-38}cm^{2}/GeV$}& \multicolumn{4}{c|}{ $neutrinos/m^{2}/GeV/10^{6}POT$} & \multicolumn{4}{c||}{$10^{-38}cm^{2}/GeV$}&\multicolumn{4}{c|}{}\\\hline
&$\Phi^{\nu}$&$\sigma_{stat}$&$\sigma_{sys}$&$\sigma_{tot}$&$\sigma^{\nu}/E$&$\sigma_{stat}$&$\sigma_{sys}$&$\sigma_{tot}$&$\Phi^{\bar{\nu}}$&$\sigma_{stat}$&$\sigma_{sys}$&$\sigma_{tot}$&$\sigma^{\bar{\nu}}/E$&$\sigma_{stat}$&$\sigma_{sys}$&$\sigma_{tot}$&\ratio{}&$\sigma_{stat}$&$\sigma_{sys}$&$\sigma_{tot}$\\\hline	
2.5 & 70.290& 1.837 & 6.446 & 6.702 & 0.746& 0.020 & 0.072 & 0.075 & 68.851& 4.589 & 6.155 & 7.678 & 0.333& 0.023 & 0.026 & 0.035 & 0.447& 0.029 & 0.017 & 0.033\\
3.5 & 78.716& 1.508 & 4.534 & 4.778 & 0.671& 0.013 & 0.036 & 0.038 & 66.833& 3.562 & 3.743 & 5.167 & 0.313& 0.017 & 0.016 & 0.024 & 0.466& 0.024 & 0.012 & 0.027\\
4.5 & 30.052& 0.624 & 1.842 & 1.945 & 0.670& 0.015 & 0.031 & 0.034 & 24.171& 1.348 & 1.472 & 1.996 & 0.308& 0.018 & 0.013 & 0.022 & 0.459& 0.026 & 0.013 & 0.029\\
6.0 & 9.557& 0.212 & 0.550 & 0.590 & 0.678& 0.016 & 0.032 & 0.036 & 6.676& 0.392 & 0.385 & 0.550 & 0.308& 0.019 & 0.013 & 0.023 & 0.455& 0.027 & 0.014 & 0.030\\
8.0 & 5.269& 0.103 & 0.258 & 0.278 & 0.697& 0.015 & 0.029 & 0.032 & 3.017& 0.160 & 0.143 & 0.214 & 0.321& 0.018 & 0.013 & 0.022 & 0.460& 0.027 & 0.010 & 0.028\\
10.5 & 3.136& 0.064 & 0.158 & 0.170 & 0.716& 0.015 & 0.029 & 0.033 & 1.625& 0.090 & 0.080 & 0.120 & 0.325& 0.019 & 0.013 & 0.023 & 0.454& 0.027 & 0.007 & 0.028\\
13.5 & 1.916& 0.034 & 0.099 & 0.104 & 0.708& 0.014 & 0.027 & 0.031 & 0.895& 0.035 & 0.044 & 0.056 & 0.324& 0.015 & 0.013 & 0.020 & 0.457& 0.022 & 0.007 & 0.023\\
16.5 & 1.173& 0.024 & 0.059 & 0.063 & 0.687& 0.015 & 0.026 & 0.030 & 0.437& 0.022 & 0.025 & 0.033 & 0.331& 0.020 & 0.013 & 0.024 & 0.482& 0.029 & 0.007 & 0.030\\
20.0 & 0.651& 0.014 & 0.034 & 0.037 & 0.698& 0.017 & 0.027 & 0.032 & 0.229& 0.014 & 0.014 & 0.019 & 0.316& 0.022 & 0.013 & 0.025 & 0.453& 0.032 & 0.006 & 0.032\\
\hline\hline
\end{tabular}	
\caption{Summary of measured quantities. Neutrino flux $\Phi^{\nu}(E)$ and antineutrino flux $\Phi^{\bar{\nu}}(E)$ and their errors  (columns 1 and 3) are in units of $neutrinos/m^{2}/GeV/10^{6}pot$. Neutrino cross section $\sigma^{\nu}(E)/E$ and antineutrino cross section $\sigma^{\bar{\nu}}(E)/E$ and their errors (columns 2 and 4) are in units of $10^{-38}cm^{2}/GeV$. Columns labeled $\sigma_{stat}$,  $\sigma_{sys}$, and $\sigma_{tot}$ give the statistical, systematic, and total errors, respectively. }	
\end{centering}
\end{sidewaystable}

\section{Acknowledgments}

This work was supported by the Fermi National Accelerator Laboratory under US Department of Energy contract No. DE-AC02-07CH11359 which included the MINERvA construction project. Construction support was also granted by the United States National Science Foundation under Award PHY-0619727 and by the University of Rochester. Support for participating scientists was provided by NSF and DOE (USA), by CAPES and CNPq (Brazil), by CoNaCyT (Mexico), by CONICYT programs including FONDECYT (Chile), by CONCYTEC, DGI-PUCP and IDI/IGI-UNI (Peru). We thank the MINOS Collaboration for use of its near detector data. We acknowledge the dedicated work of the Fermilab staff responsible for the operation and maintenance of the beamline and detector and the Fermilab Computing Division for support of data processing.
	
\bibliographystyle{apsrev4-1}
\bibliography{bibliography}
\newpage

\input{supple}

\end{document}

%% file: author.tex
\newcommand{\Rutgers}{Rutgers, The State University of New Jersey, Piscataway, New Jersey 08854, USA}
\newcommand{\Hampton}{Hampton University, Dept. of Physics, Hampton, VA 23668, USA}
\newcommand{\Dortmund}{Institute of Physics, Dortmund University, 44221, Germany }
\newcommand{\Otterbein}{Department of Physics, Otterbein University, 1 South Grove Street, Westerville, OH, 43081 USA}
\newcommand{\JMU}{James Madison University, Harrisonburg, Virginia 22807, USA}
\newcommand{\Florida}{University of Florida, Department of Physics, Gainesville, FL 32611}
\newcommand{\UCIrvine}{Department of Physics and Astronomy, University of California, Irvine, Irvine, California 92697-4575, USA}
\newcommand{\CBPF}{Centro Brasileiro de Pesquisas F\'{i}sicas, Rua Dr. Xavier Sigaud 150, Urca, Rio de Janeiro, Rio de Janeiro, 22290-180, Brazil}
\newcommand{\PUCP}{Secci\'{o}n F\'{i}sica, Departamento de Ciencias, Pontificia Universidad Cat\'{o}lica del Per\'{u}, Apartado 1761, Lima, Per\'{u}}
\newcommand{\INRM}{Institute for Nuclear Research of the Russian Academy of Sciences, 117312 Moscow, Russia}
\newcommand{\Jlab}{Jefferson Lab, 12000 Jefferson Avenue, Newport News, VA 23606, USA}
\newcommand{\Pittsburgh}{Department of Physics and Astronomy, University of Pittsburgh, Pittsburgh, Pennsylvania 15260, USA}
\newcommand{\Guanajuato}{Campus Le\'{o}n y Campus Guanajuato, Universidad de Guanajuato, Lascurain de Retana No. 5, Colonia Centro, Guanajuato 36000, Guanajuato M\'{e}xico.}
\newcommand{\Athens}{Department of Physics, University of Athens, GR-15771 Athens, Greece}
\newcommand{\Tufts}{Physics Department, Tufts University, Medford, Massachusetts 02155, USA}
\newcommand{\WM}{Department of Physics, College of William \& Mary, Williamsburg, Virginia 23187, USA}
\newcommand{\FNAL}{Fermi National Accelerator Laboratory, Batavia, Illinois 60510, USA}
\newcommand{\Purdue}{Department of Chemistry and Physics, Purdue University Calumet, Hammond, Indiana 46323, USA}
\newcommand{\MCLA}{Massachusetts College of Liberal Arts, 375 Church Street, North Adams, MA 01247}
\newcommand{\UMD}{Department of Physics, University of Minnesota -- Duluth, Duluth, Minnesota 55812, USA}
\newcommand{\Northwestern}{Northwestern University, Evanston, Illinois 60208}
\newcommand{\UNI}{Universidad Nacional de Ingenier\'{i}a, Apartado 31139, Lima, Per\'{u}}
\newcommand{\Rochester}{University of Rochester, Rochester, New York 14627 USA}
\newcommand{\Austin}{Department of Physics, University of Texas, 1 University Station, Austin, Texas 78712, USA}
\newcommand{\USM}{Departamento de F\'{i}sica, Universidad T\'{e}cnica Federico Santa Mar\'{i}a, Avenida Espa\~{n}a 1680 Casilla 110-V, Valpara\'{i}so, Chile}
\newcommand{\Geneva}{University of Geneva, 1211 Geneva 4, Switzerland}
\newcommand{\Chicago}{Enrico Fermi Institute, University of Chicago, Chicago, IL 60637 USA}
\newcommand{\hired}{}
\newcommand{\OregonState}{Department of Physics, Oregon State University, Corvallis, Oregon 97331, USA}
\newcommand{\oxford}{}
\newcommand{\umiss}{University of Mississippi, Oxford, Mississippi 38677, USA}
\newcommand{\upenn}{209 S. 33rd St. Philadelphia, PA 19104}
\newcommand{\bmeThanks}{now at SLAC National Accelerator Laboratory, Stanford, CA 94309, USA}
\newcommand{\chrismarshallThanks}{now at Lawrence Berkeley National Laboratory, Berkeley, CA 94720, USA}
\newcommand{\damartinezThanks}{now at Illinois Institute of Technology, Chicago, IL 60616, USA}
\newcommand{\mcgivernThanks}{now at Fermi National Accelerator Laboratory, Batavia, IL 60510, USA}
\newcommand{\joelmousseauThanks}{now at University of Michigan, Ann Arbor, MI 48109, USA}
\newcommand{\twaltonThanks}{now at Fermi National Accelerator Laboratory, Batavia, IL 60510, USA}
\newcommand{\jwolcottThanks}{now at Tufts University, Medford, MA 02155, USA}

\author{L.~Ren}                           \affiliation{\Pittsburgh}

\author{L.~Aliaga}                        \affiliation{\WM}  \affiliation{\PUCP}
\author{O.~Altinok}                       \affiliation{\Tufts}
\author{L.~Bellantoni}                    \affiliation{\FNAL}
\author{A.~Bercellie}                     \affiliation{\Rochester}
\author{M.~Betancourt}                    \affiliation{\FNAL}
\author{A.~Bodek}                         \affiliation{\Rochester}
\author{A.~Bravar}                        \affiliation{\Geneva}
\author{H.~Budd}                          \affiliation{\Rochester}
\author{T.~Cai}                           \affiliation{\Rochester}
\author{M.F.~Carneiro}                    \affiliation{\OregonState}
\author{H.~da~Motta}                      \affiliation{\CBPF}
\author{J.~Devan}                         \affiliation{\WM}
\author{S.A.~Dytman}                      \affiliation{\Pittsburgh}
\author{G.A.~D\'{i}az~}                   \affiliation{\Rochester}  \affiliation{\PUCP}
\author{B.~Eberly}\thanks{\bmeThanks}     \affiliation{\Pittsburgh}
\author{E.~Endress}                       \affiliation{\PUCP}
\author{J.~Felix}                         \affiliation{\Guanajuato}
\author{L.~Fields}                        \affiliation{\FNAL}  \affiliation{\Northwestern}
\author{R.~Fine}                          \affiliation{\Rochester}
\author{A.M.~Gago}                        \affiliation{\PUCP}
\author{R.Galindo}                        \affiliation{\USM}
\author{H.~Gallagher}                     \affiliation{\Tufts}
\author{A.~Ghosh}                         \affiliation{\USM}  \affiliation{\CBPF}
\author{T.~Golan}                         \affiliation{\Rochester}  \affiliation{\FNAL}
\author{R.~Gran}                          \affiliation{\UMD}
\author{J.Y.~Han}                         \affiliation{\Pittsburgh}
\author{D.A.~Harris}                      \affiliation{\FNAL}
\author{K.~Hurtado}                       \affiliation{\CBPF}  \affiliation{\UNI}
\author{M.~Kiveni}                        \affiliation{\FNAL}
\author{J.~Kleykamp}                      \affiliation{\Rochester}
\author{M.~Kordosky}                      \affiliation{\WM}
\author{T.~Le}                            \affiliation{\Tufts}  \affiliation{\Rutgers}
\author{E.~Maher}                         \affiliation{\MCLA}
\author{S.~Manly}                         \affiliation{\Rochester}
\author{W.A.~Mann}                        \affiliation{\Tufts}
\author{C.M.~Marshall}\thanks{\chrismarshallThanks}  \affiliation{\Rochester}
\author{D.A.~Martinez~Caicedo}\thanks{\damartinezThanks}  \affiliation{\CBPF}
\author{K.S.~McFarland}                   \affiliation{\Rochester}  \affiliation{\FNAL}
\author{C.L.~McGivern}\thanks{\mcgivernThanks}  \affiliation{\Pittsburgh}
\author{A.M.~McGowan}                     \affiliation{\Rochester}
\author{B.~Messerly}                      \affiliation{\Pittsburgh}
\author{J.~Miller}                        \affiliation{\USM}
\author{A.~Mislivec}                      \affiliation{\Rochester}
\author{J.G.~Morf\'{i}n}                  \affiliation{\FNAL}
\author{J.~Mousseau}\thanks{\joelmousseauThanks}  \affiliation{\Florida}
\author{D.~Naples}                        \affiliation{\Pittsburgh}
\author{J.K.~Nelson}                      \affiliation{\WM}
\author{A.~Norrick}                       \affiliation{\WM}
\author{Nuruzzaman}                       \affiliation{\Rutgers}  \affiliation{\USM}
\author{V.~Paolone}                       \affiliation{\Pittsburgh}
\author{J.~Park}                          \affiliation{\Rochester}
\author{C.E.~Patrick}                     \affiliation{\Northwestern}
\author{G.N.~Perdue}                      \affiliation{\FNAL}  \affiliation{\Rochester}
\author{M.A.~Ram\'{i}rez}                 \affiliation{\Guanajuato}
\author{R.D.~Ransome}                     \affiliation{\Rutgers}
\author{H.~Ray}                           \affiliation{\Florida}
\author{D.~Rimal}                         \affiliation{\Florida}
\author{P.A.~Rodrigues}                   \affiliation{\umiss}  \affiliation{\Rochester}
\author{D.~Ruterbories}                   \affiliation{\Rochester}
\author{H.~Schellman}                     \affiliation{\OregonState}  \affiliation{\Northwestern}
\author{C.J.~Solano~Salinas}              \affiliation{\UNI}
\author{M.~Sultana}                       \affiliation{\Rochester}
\author{S.~S\'{a}nchez~Falero}            \affiliation{\PUCP}
\author{E.~Valencia}                      \affiliation{\WM}  \affiliation{\Guanajuato}
\author{T.~Walton}\thanks{\twaltonThanks}  \affiliation{\Hampton}
\author{J.~Wolcott}\thanks{\jwolcottThanks}  \affiliation{\Rochester}
\author{M.Wospakrik}                      \affiliation{\Florida}
\author{B.~Yaeggy}                        \affiliation{\USM}

\collaboration{The MINER$\nu$A Collaboration}\ \noaffiliation
\date{\today}
\noaffiliation

%% file: supple.tex
\newpage

\onecolumngrid
\newpage
\centerline{\textbf{\Large Supplemental materials}  }

\begin{sidewaystable}
\begin{centering}
\begin{tabular}{|c|c|c c c c c c c c c|c c c c c c c c c|}
\hline
& & \multicolumn{9}{c|}{Neutrino} &\multicolumn{9}{c|}{ Antineutrino}\\\hline
&E(GeV)&2-3&3-4&4-5&5-7&7-9&9-12&12-15&15-18&18-22&2-3&3-4&4-5&5-7&7-9&9-12&12-15&15-18&18-22\\\hline
&&5.67&1.67&1.35&1.35&1.13&1.03&0.83&0.78&0.81&1.31&0.70&0.50&0.45&0.45&0.45&0.41&0.45&0.41\\
&2-3& &1.50&0.88&0.84&0.80&0.78&0.72&0.63&0.64&0.67&0.47&0.37&0.34&0.35&0.36&0.34&0.33&0.31\\
&3-4& & &1.19&0.81&0.74&0.71&0.65&0.66&0.69&0.54&0.39&0.34&0.32&0.32&0.32&0.31&0.34&0.32\\
&5-7& & & &1.26&0.77&0.76&0.65&0.68&0.70&0.54&0.37&0.33&0.35&0.33&0.33&0.32&0.35&0.33\\
Neutrino&7-9& & & & &1.04&0.76&0.70&0.66&0.66&0.47&0.36&0.32&0.32&0.34&0.34&0.33&0.33&0.30\\
&9-12& & & & & &1.06&0.73&0.66&0.65&0.45&0.36&0.32&0.33&0.35&0.37&0.34&0.32&0.30\\
&12-15& & & & & & &0.93&0.63&0.62&0.38&0.34&0.31&0.30&0.33&0.34&0.33&0.30&0.29\\
&15-18& & & & & & & &0.89&0.68&0.33&0.29&0.29&0.30&0.29&0.30&0.29&0.32&0.31\\
&18-22& & & & & & & & &1.02&0.34&0.29&0.30&0.30&0.29&0.28&0.29&0.34&0.33\\\hline
&2-3& & & & & & & & & &1.20&0.30&0.21&0.20&0.20&0.19&0.19&0.18&0.17\\
&3-4& & & & & & & & & & &0.55&0.17&0.16&0.17&0.17&0.16&0.15&0.14\\
&4-5& & & & & & & & & & & &0.48&0.15&0.15&0.14&0.15&0.14&0.14\\
&5-7& & & & & & & & & & & & &0.52&0.14&0.14&0.14&0.15&0.14\\
Antineutrino&7-9& & & & & & & & & & & & & &0.51&0.16&0.15&0.14&0.13\\
&9-12& & & & & & & & & & & & & & &0.54&0.16&0.14&0.13\\
&12-15& & & & & & & & & & & & & & & &0.39&0.14&0.14\\
&15-18& & & & & & & & & & & & & & & & &0.56&0.15\\
&18-22& & & & & & & & & & & & & & & & & &0.65\\
\hline
\end{tabular}
\caption{Covariance matrix corresponding to total error for the extracted neutrino cross section in the FHC and antineutrino cross section in the RHC beam mode. The covariance elements are in units of $(\sigma/E)^{2}$, which is $(10^{-38} cm^{2} / \textrm{GeV} )^{2}$, and scaled by a factor of 1000. }
\label{app:xsec_cov}
\end{centering}
\end{sidewaystable}

\begin{table}[ht]
\begin{centering}
\begin{tabular}{|c|c c c c c c c c c|}
\hline
E(GeV)&2-3&3-4&4-5&5-7&7-9&9-12&12-15&15-18&18-22\\\hline
2-3&1.080&0.088&0.021&0.002&-0.024&-0.018&-0.030&-0.019&-0.010\\
3-4& &0.718&0.119&0.106&0.056&0.039&0.010&0.015&0.019\\
4-5& & &0.816&0.150&0.103&0.065&0.032&0.031&0.024\\
5-7& & & &0.919&0.107&0.083&0.046&0.045&0.038\\
7-9& & & & &0.796&0.060&0.033&0.033&0.030\\
9-12& & & & & &0.788&0.030&0.031&0.027\\
12-15& & & & & & &0.507&0.031&0.015\\
15-18& & & & & & & &0.872&0.017\\
18-22& & & & & & & & &1.053\\
\hline
\end{tabular}
\caption{Covariance matrix of extracted cross section ratio, \ratio{}, scaled by 1000. }
\label{app:r_cov}
\end{centering}
\end{table}

\begin{table}[ht]
\begin{centering}
\begin{tabular}{|c|c c c c c c c c c|}
\hline
E(GeV)&2-3&3-4&4-5&5-7&7-9&9-12&12-15&15-18&18-22\\\hline
2-3&44.869&21.466&-0.026&0.475&0.832&0.629&0.421&0.109&0.017\\
3-4& &22.823&3.814&1.333&0.847&0.548&0.337&0.160&0.075\\
4-5& & &3.784&0.815&0.295&0.152&0.078&0.076&0.050\\
5-7& & & &0.348&0.099&0.055&0.029&0.023&0.015\\
7-9& & & & &0.077&0.032&0.019&0.011&0.006\\
9-12& & & & & &0.029&0.012&0.006&0.003\\
12-15& & & & & & &0.011&0.004&0.002\\
15-18& & & & & & & &0.004&0.002\\
18-22& & & & & & & & &0.001\\
\hline
\end{tabular}
\caption{Covariance matrix for the extracted neutrino flux in the FHC beam mode. The covariance elements are in units of  $(\nu_{\mu}/m^{2}/\textrm{GeV}/10^{6} \textrm{POT})^{2}$.}
\end{centering}
\end{table}

\begin{table}[ht]
\begin{centering}
\begin{tabular}{|c|c c c c c c c c c|}
\hline
E(GeV)&2-3&3-4&4-5&5-7&7-9&9-12&12-15&15-18&18-22\\\hline
2-3&58.944&17.011&-0.415&0.032&0.386&0.250&0.137&-0.003&-0.004\\
3-4& &26.699&2.211&0.623&0.370&0.216&0.117&0.037&0.020\\
4-5& & &4.166&0.488&0.140&0.069&0.038&0.030&0.017\\
5-7& & & &0.302&0.038&0.019&0.010&0.008&0.004\\
7-9& & & & &0.046&0.009&0.005&0.002&0.001\\
9-12& & & & & &0.014&0.003&0.001&0.001\\
12-15& & & & & & &0.003&0.001&0.000\\
15-18& & & & & & & &0.001&0.000\\
18-22& & & & & & & & &0.000\\
\hline
\end{tabular}
\caption{Covariance matrix for the extracted antineutrino flux in the RHC beam mode.  The covariance elements are in units of  $(\nu_{\mu}/m^{2}/\textrm{GeV}/10^{6} \textrm{POT})^{2}$.}
\end{centering}
\end{table}